\documentstyle[12pt,axodraw,epsfig]{article}

\newcommand{\imag}{\Im {\rm m}}
\newcommand{\real}{\Re {\rm e}}

\def\lsim{\:\raisebox{-0.5ex}{$\stackrel{\textstyle<}{\sim}$}\:}
\def\gsim{\:\raisebox{-0.5ex}{$\stackrel{\textstyle>}{\sim}$}\:}

\begin{document}

\renewcommand{\thefootnote}{\fnsymbol{footnote}}

\mbox{ } \\[-1cm]
\mbox{ }\hfill KEK--TH--753\\
\mbox{ }\hfill TUM--HEP--401/01\\
\mbox{ }\hfill hep--ph/0103284\\
\mbox{ }\hfill \today\\

\begin{center}
{\Large\bf CP Violation in Tau Slepton Pair Production \\[2mm]
           at Muon Colliders}\\[1cm] 
S.Y. Choi$^1$,\, M. Drees$^2$,\, B. Gaissmaier$^2$\, and Jae Sik Lee$^3$
\end{center}

\bigskip 

\begin{center}
$^1${\it Department of Physics, Chonbuk National University, 
     Chonju 561--756, Korea} \\[1mm]
$^2${\it Physik Dept., TU M\"{u}nchen, James Franck Str., D--85748 
     Garching, Germany} \\[1mm]
$^3${\it KEK Theory Group, Tsukuba, Ibaraki 305--0801, Japan}
\end{center}

\bigskip
\bigskip 
\bigskip 

\begin{abstract}
We discuss in detail signals for CP violation in the Higgs boson and
tau--slepton sectors through the production processes $\mu^+\mu^-
\rightarrow \tilde{\tau}_i^- \tilde{\tau}_j^+$, where $i,j=1,2$ label
the two $\tau$ slepton mass eigenstates in the minimal supersymmetric
standard model. We assume that the soft breaking parameters of third
generation sfermions contain CP violating phases, which induce CP
violation in the Higgs sector through quantum corrections. We
classify all the observables for probing CP violation in the Higgs
boson and $\tau$ slepton sectors. These observables depend on the initial
muon beam polarization, where we include transverse polarization
states. If the heavy Higgs bosons can decay into tau slepton pairs, a
complete determination of the CP properties of the neutral Higgs boson
and $\tau$--slepton systems is possible. The interference between the
Higgs boson and gauge boson contributions could also provide a
powerful method for probing CP violation, if transversely polarized
muon beams are available. We show in detail how to directly measure
CP violation in the tau slepton system, under the assumption that the
neutral Higgs mixing angles are determined through the on--shell
production of the neutral Higgs bosons.
\vskip 0.5cm

\noindent
PACS number(s): 11.30.Pb, 11.30.Er
\end{abstract}

\newpage

\section{Introduction}
\label{sec:introduction}
\setcounter{footnote}{1}

CP violation was observed in the neutral kaon system \cite{Kaon} and
it is strongly suggested by recent experiments in $B$--meson decays
\cite{B-meson}. In addition, CP violation constitutes one of the
conditions for a dynamical generation of the cosmological baryon
asymmetry \cite{sakharov}.  In the Standard Model (SM), which contains
only one physical neutral Higgs boson, the only source of CP violation
is the complex phase of the quark mixing matrix \cite{km}.\footnote{In
principle CP could also be violated in the SM by the QCD
$\theta-$term, but bounds on the electric dipole moment of the neutron
force $\theta_{\rm QCD}$ to be very small $\lsim 10^{-10}$.} On the
other hand, models with multiple Higgs doublets can have additional
sources of CP violation, e.g. neutral and/or charged Higgs bosons of
no definite CP quantum number.

Supersymmetry (SUSY) is now widely regarded to be the most plausible
extension of the SM; among other things, it stabilizes the gauge
hierarchy \cite{6} and allows for the Grand Unification of all known
gauge interactions \cite{7}. Of course, supersymmetry must be (softly)
broken to be phenomenologically viable. In general this introduces a
large number of unknown parameters, many of which can be complex
\cite{DS}. CP--violating phases associated with sfermions of the first
and, to a lesser extent, second generation are severely constrained by
bounds on the electric dipole moments of the electron, neutron and
muon. However, it has recently been realized \cite{9} that
cancellations between different diagrams allow some combinations of
these phases to be quite large. Even in models with universal boundary
conditions for soft breaking masses at some very high energy scale,
the relative phase between the supersymmetric higgsino mass parameter
$\mu$ and the universal trilinear soft breaking parameter $A_0$ can be
${\cal O}(1)$ \cite{10}.  If universality is not assumed, the phases
of third generation trilinear soft breaking parameters are essentially
unconstrained. In fact, some of these phases might be large
\cite{worah} so as to provide non--SM sources of CP violation required
for a dynamical generation of the baryon asymmetry of the Universe.

Recently it has been realized that the explicit CP violation in the
mass matrices of the third generation squarks with such possibly large
CP phases\footnote{If $\tan\beta$ is large, the third--generation
CP phases are constrained by the electric dipole moments of the
electron and neutron because in this case their contributions at the
two--loop level could be sizable \cite{CKP}.} can induce sizable CP
violation in the MSSM Higgs sector through loop corrections
\cite{EXCP1, EXCP2, EXCP3, CDL, EXCP4}. This induced CP violation in
the MSSM Higgs sector can affect the phenomenology of the Higgs bosons
at present and future colliders \cite{EXCP1, EXCP2, EXCP4, CD2,
EXCP_FC, SL1, SL2, eri}. These phases also directly affect the
couplings of Higgs bosons to third generations sfermions. These
couplings play an important role in the calculation of loop-induced
couplings of Higgs bosons to photons and gluons as well as in the
production of third generation sfermions at $\mu^+ \mu^-$ colliders
\cite{CD2}.

In the past few years a considerable amount of effort has been devoted
to investigations of the physics potential of high energy $\mu^+\mu^-$
colliders \cite{MUCOL} as a plausible future experimental
program. Since muons emit far less synchrotron radiation than
electrons do, a muon collider might be significantly smaller and
cheaper than an $e^+ e^-$ collider operating at the same
center--of--mass energy $\sqrt{s}$. The main physics advantage of muon
colliders compared to the conventional $e^+e^-$ and hadron colliders
is that the larger Yukawa coupling of muons in many cases admits
copious production of Higgs bosons as $s-$channel resonances, allowing
to perform precision measurements of their properties
\cite{MUCOL,S_H,SONI,CD2,GGP,bartl}. In particular, they can allow us
to search for CP--violation in the couplings of Higgs bosons to not
only heavy SM fermions but also to third generation sfermions, which
have large Yukawa couplings.

It is difficult to probe these CP phases through processes controlled
by gauge interactions, where large CP--odd asymmetries can emerge only
if some sfermion mass eigenstates are closely degenerate, with mass
splitting of the order of the decay width, in which case flavor or
chirality oscillations can occur \cite{11a,11b,CD1}. On the other
hand, in the MSSM CP--violating phases can appear at tree level in
the couplings of a single sfermion species to neutral Higgs
bosons. These phases can give rise to large CP--odd asymmetries
regardless of sfermion mass splittings.  Here we focus on
$\tilde{\tau}$ pair production. Unlike sfermions of the first two
generations, $\tilde{\tau}$'s generally have sizable couplings to
heavy Higgs bosons even if the latter are much heavier than
$m_Z$. Furthermore, unlike for $\tilde{b}$ and $\tilde {t}$ production
the charge of a produced $\tilde{\tau}$ is usually readily measurable;
this is necessary for the construction of most CP--odd
asymmetries. Finally, in most models sleptons are significantly
lighter than squarks, making it easier to study them at lepton
colliders.

In the present work we extend the previous work \cite{CD2}
significantly in order to discuss in more detail signals for CP
violation in the production processes $\mu^-\mu^+ \rightarrow
\tilde{\tau}_i^- \tilde{\tau}_j^+$ where $i,j=1,2$ label the two
$\tau$ slepton mass eigenstates. We work in the framework of the MSSM
with explicit CP violation. More specifically, we present a general
formalism and a detailed analysis of the effects of the CP--violating
Higgs boson mixing, and of the CP phases in the tau slepton mass
matrix, on the polarized cross section for tau--slepton pair
production. We consider both longitudinal and transverse polarization
of the initial muon and anti--muon beams. A detailed procedure is
suggested for measuring CP violation directly in the tau slepton
system, under the assumption that the neutral Higgs mixing angles are
determined through the on--shell production of the neutral Higgs
bosons as demonstrated in Ref.~\cite{eri}. The polarization
observables allow for the complete determination of the CP properties
of the neutral Higgs boson and $\tau$ slepton systems. The
interference between Higgs and gauge boson exchange contributions
plays a crucial role in this analysis.

The remainder of this article is organized as follows.
Section~\ref{sec:susy mixing} is devoted to a brief review of the mass
spectra and mixing patterns of the neutral Higgs bosons and tau
sleptons, focusing on the role of the CP phases \cite{EXCP2, CDL,
EXCP4}. In Sec.~\ref{sec:production} we present the helicity
amplitudes of the production of a $\tau$--slepton pair in $\mu^+\mu^-$
collisions with polarized muon beams.  We also give a complete
classification of the CP and CP$\tilde{\rm T}$ transformation
properties of the polarization observables. In
Sec.~\ref{sec:reconstruction} we show how to extract the rate and
polarization asymmetries by taking appropriate configurations of the
muon and anti--muon spins. We then perform a detailed numerical
analysis for a representative SUSY parameter set so as to get a
concrete estimate of the relative usefulness of those observables.
Section~\ref{sec:conclusion} is devoted to a brief summary of our
findings and to conclusions.

\section{Supersymmetric Particle Mixing}
\label{sec:susy mixing}
\setcounter{footnote}{1}

Bounds on CP violating flavor changing processes tell us that the CP
violating phases associated with flavor off--diagonal trilinear scalar
interactions must be strongly suppressed \cite{italians}. We therefore
neglect all these flavor changing CP violating phases in the present
work, so that the scalar soft mass matrices and trilinear parameters
are flavor diagonal and the complex trilinear terms are proportional
to the corresponding fermion Yukawa couplings. Clearly the Yukawa
interactions of the top and bottom (s)quarks play the most significant
role in radiative corrections to the Higgs sector. Furthermore, for
large values of $\tan\beta$ the $\tilde\tau_L$ and $\tilde\tau_R$
states are expected to mix strongly.

\subsection{CP--violating neutral Higgs boson mixing}

In this section, we give a brief review of the calculation 
\cite{CDL} of the Higgs boson mass matrix based on the full one--loop 
effective potential, valid for all values of the relevant 
third--generation squark soft--breaking parameters. The CP--violating 
phases in the top and bottom squark sectors cause scalar--pseudoscalar
mixing at one--loop level.

The MSSM contains two Higgs doublets $H_1, \ H_2$, with hypercharges
$Y(H_1) = -Y(H_2) = -1/2$. Here we are only interested in the neutral
components, which we write as
\begin{eqnarray}
\label{e1}
H_1^0=\frac{1}{\sqrt{2}}\left( \phi_1 + i a_1 \right); \ \ \ \ \
H_2^0=\frac{{\rm e}^{i \xi}}{\sqrt{2}}\left( \phi_2 + i a_2 \right),
\end{eqnarray}
where $\phi_{1,2}$ and $a_{1,2}$ are real fields. The constant phase
$\xi$ can be set to zero at tree level, but will in general become
non--zero once loop corrections are included. 

The mass matrix of the neutral Higgs bosons can be computed from the
effective potential \cite{CW}
\begin{eqnarray}
\label{e2}
V_{\rm Higgs}\hskip -0.3cm 
   &= \, \frac{1}{2}m_1^2\left(\phi_1^2+a_1^2\right)
     +\frac{1}{2}m_2^2\left(\phi_2^2+a_2^2\right)
     -\left|m_{12}^2\right|\left(\phi_1\phi_2-a_1 a_2\right) 
           \cos (\xi + \theta_{12}) \nonumber \\ 
   & +\left|m^2_{12}\right|
      \left(\phi_1 a_2 +\phi_2 a_1\right)\sin(\xi+\theta_{12})
     +\frac{\hat{g}^2}{8} {\cal D}^2 
     +\frac{1}{64\,\pi^2} {\rm Str} \left[
           {\cal M}^4 \left(\log\frac{{\cal M}^2}{Q^2} 
	                  - \frac{3}{2}\right)\right].
\end{eqnarray}
We have allowed the soft breaking parameter $m^2_{12} = \left|
m^2_{12} \right| {\rm e}^{i \theta_{12}}$ to be complex, and we have
introduced the quantities
\begin{eqnarray}
\label{e5}
{\cal D} = \phi_2^2 + a_2^2 - \phi_1^2 - a_1^2; \ \ \ \ \ \hat{g}^2 
         = \frac{g^2 + g'^2}{4},
\end{eqnarray}
where the symbols $g$ and $g'$ stand for the SU(2)$_L$ and U(1)$_Y$
gauge couplings, respectively. $Q$ in Eq.~(\ref{e2}) is the
renormalization scale; the parameters of the tree--level potential, in
particular the mass parameters $m_1^2, \ m_2^2 $ and $m_{12}^2$, are
running parameters, taken at scale $Q$. The potential (\ref{e2}) is
then independent of $Q$, up to two--loop corrections.

The matrix ${\cal M}$ in Eq.~(\ref{e2}) is the field--dependent mass 
matrix of  all modes that couple to the Higgs bosons. 
The by far dominant contributions come from the third generation 
quarks and squarks. The (real) masses of the former are given by
\begin{eqnarray}
\label{e3}
m_b^2 = \frac{1}{2}\, |h_b|^2 \left( \phi_1^2 + a_1^2
\right); \ \ \ \
m_t^2 = \frac{1}{2}\, |h_t|^2 \left( \phi_2^2 + a_2^2
\right), 
\end{eqnarray}
where $h_b$ and $h_t$ are the bottom and top Yukawa couplings. The
corresponding squark mass matrices can be written as
\begin{eqnarray}
&& {\cal M}_{\tilde t}^2= \left(\begin{array}{cc} 
       m^2_{\widetilde Q}+m_t^2-\frac{1}{8}
       \left(g^2-\frac{g'^2}{3}\right){\cal D}
  & - h_t^*\left[\,A_t^*\left(H_2^0 \right)^* + \mu H_1^0\right]\\[2mm]
    - h_t  \left[\,A_t H^0_2+\mu^* \left( H_1^0 \right)^*\right] 
  & m^2_{\widetilde U} + m_t^2 - \frac{g'^2}{6} {\cal D}
                             \end{array} \right),\nonumber\\[3mm]
&& {\cal M}_{\tilde b}^2= \left(\begin{array}{cc} 
       m^2_{\widetilde Q}+m_b^2+\frac{1}{8}
       \left(g^2+\frac{g'^2}{3}\right){\cal D}
 & - h_b^*\left[\,A_b^*\left(H_1^0 \right)^* + \mu H_2^0\right]\\[2mm]
   - h_b  \left[\,A_b H_1^0 + \mu^* \left(H_2^0 \right)^*\right] 
 & m^2_{\widetilde D} + m_b^2 + \frac{g'^2}{12} {\cal D}
                           \end{array} \right). 
\label{e4}
\end{eqnarray}
Here, $H_1^0$ and $H_2^0$ are given by Eq.~(\ref{e1}) while $m_t^2$
and $m_b^2$ are as in Eq.~(\ref{e3}) and ${\cal D}$ has been defined
in Eq.~(\ref{e5}). In Eq.~(\ref{e4}) $m^2_{\widetilde Q}, \ 
m^2_{\widetilde U}$ and $m^2_{\widetilde D}$ are real soft breaking 
parameters, $A_b$ and $A_t$ are complex soft breaking parameters, and 
$\mu$ is the complex supersymmetric Higgs(ino) mass parameter.  

The mass matrix of the neutral Higgs bosons can now be computed from
the matrix of second derivatives of the potential (\ref{e2}), where
$m_1^2, \ m_2^2$ and $m_{12}^2 \sin( \xi + \theta_{12})$ are
determined by the stationarity conditions\footnote{The condition
$\partial V / \partial a_2 = 0$ gives the same constraint as $\partial
V / \partial a_1 = 0$ \cite{CDL}.} $\partial V/ \partial \phi_1 =
\partial V / \partial \phi_2 = \partial V / \partial a_1 = 0$. The
massless state $G^0 = a_1 \cos \beta - a_2 \sin \beta$ is the
would--be Goldstone mode ``eaten'' by the longitudinal $Z$ boson. We
are thus left with a squared mass matrix ${\cal M}_H^2$ for the three
states $a = a_1 \sin \beta + a_2 \cos \beta, \ \phi_1$ and
$\phi_2$. This matrix is real and symmetric, i.e. it has 6 independent
entries. CP violation is caused by the appearance of nonvanishing $a
\phi_1$ or $a \phi_2$ entries of this mass matrix. The sizes of these
entries are controlled by the quantities
\begin{eqnarray}
\label{e9}
\Delta_{\tilde t}=\frac{\imag(A_t\,\mu\,{\rm e}^{i\xi})}
                  {m^2_{\tilde{t}_2}-m^2_{\tilde{t}_1}}\,, \qquad
\Delta_{\tilde b}=\frac{\imag(A_b\,\mu\,{\rm e}^{i\xi})}
                  {m^2_{\tilde{b}_2}-m^2_{\tilde{b}_1}}.
\end{eqnarray}
Explicit expressions for ${\cal M}_H^2$ can be found in
Ref.~\cite{CDL}. 

Since ${\cal M}_H^2$ is real and symmetric, it can be diagonalized with
a three--dimensional orthogonal rotation matrix $O$;
\begin{eqnarray} \label{eq:hrot}
\left(\begin{array}{c}
       a      \\
       \phi_1 \\
       \phi_2
      \end{array}\right)\,=\,O\,
\left(\begin{array}{c}
       H_1 \\
       H_2 \\
       H_3
      \end{array}\right)\, .
\end{eqnarray}
Our convention for the three mass eigenvalues is $m_{H_1}\leq
m_{H_2}\leq m_{H_3}$. The loop--corrected neutral Higgs boson sector
is thus determined by fixing the values of the following parameters:
$m_A$, which becomes the mass of the CP--odd Higgs boson if CP is
conserved, and $\tan\beta$ fix the tree--level Higgs potential; and
$\mu$, $A_t$, $A_b$ and the soft--breaking third generation sfermion
masses $m_{\tilde Q}$, $m_{\tilde U}$, and $m_{\tilde D}$, which fix
the third generation squark mass matrices. After minimization of the
potential the rephasing invariant sum $\theta_{12} + \xi$ of the
radiatively induced phase $\xi$ and the phase $\theta_{12}$ of the
soft breaking parameter $m^2_{12}$ is no longer an independent
parameter.\footnote{As discussed in \cite{CDL}, $\xi$ and
$\theta_{12}$ are not separately physical parameters. For example, one
or the other can be set to zero in certain phase conventions for the
fields. Similar remarks hold for the phases of $A_t$, $A_b$ and
$\mu$. Altogther there are only three rephasing invariant
(i.e. physical) phases, which we write as $\theta_{12}+\xi$, $\arg(A_t
\mu {\rm e}^{i\xi})$ and $\arg(A_b \mu {\rm e}^{i\xi})$. The
minimization of the potential fixes one of these combinations, leaving
two independent physical phases as free input parameters.} The
physically meaningful CP phases in the Higgs sector are thus the
phases of the re--phasing invariant combinations $A_t\, \mu \, {\rm
e}^{i\xi}$ and $A_b\, \mu \, {\rm e}^{i\xi}$ appearing in
Eqs.~(\ref{e9}). We refer to Ref.\cite{CDL} for further details on
neutral Higgs boson mixing in the presence of CP--violation. This
mixing changes all the couplings of the Higgs fields so that the
effects of CP violation in the Higgs sector can be probed through
various processes \cite{EXCP_FC,SL1,SL2,eri}.

\subsection{Tau slepton mixing}

The mass matrix squared ${\cal M}^2_{\tilde{\tau}}$ for the 
left-/right-handed tau sleptons is given by
\begin{eqnarray}
{\cal M}^{\,2}_{\tilde{\tau}}=\left(\begin{array}{cc}
          X_\tau             & Z_\tau\, {\rm e}^{-i\phi_\tau}\\[2mm]
Z_\tau\, {\rm e}^{i\phi_\tau}&       Y_\tau 
                           \end{array}\right)\,,
\end{eqnarray}
where the matrix elements are defined as
\begin{eqnarray} \label{eq:xyztau}
X_\tau&=&m^2_{\tilde{\tau}_{_L}} +m^2_\tau
       +\frac{1}{2}(m^2_Z-2m^2_W)\cos 2\beta\,,\nonumber\\
Y_\tau&=& m^2_{\tilde{\tau}_{_R}} +m^2_\tau
       +(m^2_W-m^2_Z)\cos 2\beta\,, \nonumber\\
Z_\tau&=&m_\tau\,|\,A^*_\tau + \mu\, {\rm e}^{i\xi}\tan\beta\,|\,,
        \nonumber\\
\phi_\tau&=& {\rm arg}(-A_\tau - \mu^*\, {\rm e}^{-i\xi}\tan\beta)\,,
\end{eqnarray}
where $m^2_{\tilde{\tau}_{_{L,R}}}$ are the left/right--handed 
soft--SUSY--breaking tau slepton masses squared, respectively.

The mass matrix squared ${\cal M}^{\,2}_{\tilde{\tau}}$ can be 
diagonalized by a unitary transformation $U_\tau$\,:
\begin{eqnarray}
U^\dagger_\tau\,{\cal M}^2_{\tilde\tau} \, U_\tau
  ={\rm diag}(\, m^2_{\tilde{\tau}_1}, m^2_{\tilde{\tau}_2})\,,
\end{eqnarray}
with the mass ordering $m_{\tilde{\tau}_1}\leq m_{\tilde{\tau}_2}$ as
a convention. The diagonalization matrix can be parameterized as
\begin{eqnarray}
U_\tau =\left(\begin{array}{cc}
 \cos\theta_\tau   & -\sin\theta_\tau\, {\rm e}^{-i\phi_\tau} \\[2mm]
 \sin\theta_\tau\, {\rm e}^{i\phi_\tau} &  \cos\theta_\tau
           \end{array}\right)\,,
\end{eqnarray}
taking the ranges, $-\pi/2\leq \theta_\tau \leq 0$ and 
$0\leq\phi_\tau\leq 2\pi$. The tau slepton mass eigenvalues 
and mixing angles are then given by
\begin{eqnarray}
&& m^2_{\tilde{\tau}_{1,2}}=\overline{M}^2_{\tilde\tau} \mp
        \frac{\Delta_\tau}{2} 
        \,,\nonumber\\[1mm]
&& \sin2\theta_\tau =-2\,\frac{Z_\tau}{\Delta_\tau},\quad
   \cos2\theta_\tau =-\frac{X_\tau-Y_\tau}{\Delta_\tau}.
\end{eqnarray}
The phenomenological parameters $\overline{M}^2_{\tilde\tau}$ and
$\Delta_\tau$ are related to the tau slepton masses as
\begin{eqnarray}
&& \overline{M}^2_{\tilde\tau}\equiv\frac{m^2_{\tilde{\tau}_2}
                      +m^2_{\tilde{\tau}_1}}{2}
                 = \frac{X_\tau+Y_\tau}{2}\,, 
   \nonumber\\
&& \Delta_\tau\equiv m^2_{\tilde{\tau}_2}-m^2_{\tilde{\tau}_1}
                 = \sqrt{(X_\tau-Y_\tau)^2 + 4Z^2_\tau}\,.
\end{eqnarray}
Clearly tau slepton left--right mixing is enhanced for large
$\tan\beta$ and large $|\mu|$.

\section{Tau Slepton Pair Production}
\label{sec:production}
\setcounter{footnote}{1}

In this section, we present all the Feynman rules needed for the
production process $\mu^+\mu^-\rightarrow\tilde{\tau}_i^-
\tilde{\tau}_j^+ $ $(i,j=1,2$), derive the production amplitudes, and
describe how to obtain the polarized cross sections with initial muon
and anti--muon polarizations.  We then classify all the polarization
and rate asymmetries according to their CP and CP$\tilde{\rm T}$
transformation properties, and discuss the CP$\tilde{\rm T}$--even and
--odd combinations of the neutral Higgs boson propagators, neglecting
the $Z$ boson width.

\subsection{Feynman rules}
\label{subsec:Feynman rules}

The couplings of the $\gamma$ and the neutral gauge boson $Z$ to
fermions in the MSSM is described by the same Lagrangian as in the SM:
\begin{eqnarray}
{\cal L}_{Vff}=- eQ_f \bar{f}\gamma_\mu f\,A^\mu-\frac{g}{c_W}
                    \bar{f}\,\gamma_\mu \left[\,(T_{f3}-Q_f\,s^2_W)P_-
                                   -Q_f\,s^2_WP_+ \right]f\, Z^\mu\,,
\end{eqnarray}
with\, $e=g s_W$ and the chirality projection operators $P_\pm=(1\pm
\gamma_5)/2$. $Q_f$ is the electric charge of fermion $f$ in units of
the proton charge. The couplings of the neutral Higgs bosons to
leptons and quarks are described by the Lagrangian
\begin{eqnarray}
{\cal L}_{Hff}&=& -\frac{h_l}{\sqrt{2}}\sum_{k=1}^3 
    \bar{\ell}\left[\,O_{2k}-is_\beta\,O_{1k}\gamma_5\right]\ell H_k
    -\frac{h_d}{\sqrt{2}}\sum_{k=1}^3 
  \bar{d}\left[\,O_{2k}-is_\beta\,O_{1k}\gamma_5\right] d H_k\nonumber\\
   &&  -\frac{h_u}{\sqrt{2}}\sum_{k=1}^3
       \bar{u}\left[\,O_{3k}- ic_\beta\,O_{1k}\gamma_5\right]u H_k\, .
\end{eqnarray}
Here $h_l$, $h_d$, and $h_u$ are the lepton and quark
Yukawa couplings:
\begin{eqnarray}
h_l=\frac{g\,m_l}{\sqrt{2}m_W c_\beta},\quad
h_d=\frac{g\,m_d}{\sqrt{2}m_W c_\beta},\quad
h_u=\frac{g\,m_u}{\sqrt{2}m_W s_\beta}\,,
\end{eqnarray}
respectively, with $c_\beta \equiv \cos\beta$ and $s_\beta \equiv
\sin\beta$. It is then clear that all the neutral Higgs bosons couple
dominantly to the third generation fermions $t$, $b$ and $\tau$ and
they couple to a muon about 200 times more strongly than to an
electron -- the {\it primary reason} for having a muon collider.

The couplings of the $\gamma$ and the neutral gauge boson $Z$ to
tau sleptons in the MSSM are described in the mass eigenstate 
basis by the Lagrangian
\begin{eqnarray}
{\cal L}_{V\tilde{\tau}_i\tilde{\tau}_j}
  = i\,e A^\mu \delta_{ij}\,(\tilde{\tau}^*_i
     \stackrel{\longleftrightarrow}{\partial_\mu}\tilde{\tau}_j)
   -i\, \frac{g}{c_W}\,Z^\mu\,Q^Z_{ij}\,
     (\tilde{\tau}^*_i\stackrel{\longleftrightarrow}{\partial_\mu}
      \tilde{\tau}_j)\,,
\end{eqnarray}
where $Q^Z_{ij}$ are expressed in terms of the tau slepton mixing 
matrix $U_\tau$ as
\begin{eqnarray} \label{eq:qz}
Q^Z_{ij}=s^2_W\,\delta_{ij}-\frac{1}{2}(U_\tau)^*_{1i}(U_\tau)_{1j}\,.
\end{eqnarray}
On the other hand, the Feynman rules for the Higgs boson couplings
to tau slepton pairs, involving all the mixing and phases and 
including the phase rotations of the scalar tau fields, can be 
written as
\begin{eqnarray}
{\cal L}_{H\tilde{\tau}\tilde{\tau}}=
  -\frac{gm_Z}{c_W}\, O_{\alpha k}
  (U_\tau)^*_{\beta i}(U_\tau)_{\gamma j}
  V_{\alpha;\beta\gamma} \, H_k \tilde{\tau}^*_i\tilde{\tau}_j
  \equiv -\frac{g m_Z}{c_W}\, V_{k;ij} \, H_k 
  \tilde{\tau}^*_i\tilde{\tau}_j
\end{eqnarray}
where $\alpha$ labels the three neutral Higgs boson interaction eigenstates
$\{a,\phi_1,\phi_2\}$, and $\{\beta,\gamma\}$ denote the chiralities 
$\{L,R\}$ of the $\tilde\tau$ interaction eigenstates.
The chiral couplings $V_{\alpha;\beta\gamma}$ 
for the scalar tau leptons can be obtained in a rather tedious but 
straightforward way as 
\begin{eqnarray} \label{eq:hstaucoup}
&& V_a=\frac{m_\tau}{2m^2_Z}
       \left(\begin{array}{cc}
         0   & i(A^*_\tau \tan\beta-\mu {\rm e}^{i\xi}) \\[2mm]
        -i(A_\tau \tan\beta-\mu^* {\rm e}^{-i\xi}) & 0
	     \end{array}\right)\,,\nonumber\\[3mm]
&& V_{\phi_1}=c_\beta
       \left(\begin{array}{cc}
         s^2_W-\frac{1}{2}+\frac{m^2_\tau}{m^2_Zc^2_\beta} &
	 -\frac{m_\tau A^*_\tau}{2m^2_Z c^2_\beta} \\[2mm]
	 -\frac{m_\tau A_\tau}{2m^2_Z c^2_\beta} & 
         -s^2_W+\frac{m^2_\tau}{m^2_Zc^2_\beta}
	     \end{array}\right)\,,\nonumber\\[3mm]
&& V_{\phi_2}=-s_\beta
       \left(\begin{array}{cc}
         s^2_W-\frac{1}{2} &
 \frac{m_\tau \mu{\rm e}^{i\xi}}{2m^2_Z c_\beta s_\beta} \\[2mm]
 \frac{m_\tau \mu^*{\rm e}^{-i\xi}}{2m^2_Z c_\beta s_\beta} &
         -s^2_W
	     \end{array}\right)\, .
\end{eqnarray}
These $2\times 2$ matrices determine the dimensionless couplings
$V_{k;ij}$ through the relation:
\begin{eqnarray}
V_{k;ij}= O_{\alpha k}\,(U_\tau)^*_{\beta i}\,
          (U_\tau)_{\gamma j}\, V_{\alpha;\beta\gamma}.
\end{eqnarray}
\vspace*{0cm} 
\begin{center}
\begin{picture}(300,100)(0,0)

\Text(5,85)[]{$\mu^-$}
\ArrowLine(10,75)(35,50)
\ArrowLine(35,50)(10,25)
\Text(5,15)[]{$\mu^+$}

\DashLine(35,50)(75,50){4}
\Text(55,37)[]{$H_k$}

\DashLine(75,50)(100,75){4}
\Text(110,85)[]{$\tilde{\tau}^-_i$}
\DashLine(100,25)(75,50){4}
\Text(110,15)[]{$\tilde{\tau}^+_j$}

\Text(185,85)[]{$\mu^-$}
\ArrowLine(190,75)(215,50)
\Text(185,15)[]{$\mu^+$}
\ArrowLine(215,50)(190,25)

\Photon(215,50)(255,50){4}{8}
\Text(235,37)[]{$\gamma,\,Z$}

\DashLine(255,50)(280,75){4}
\Text(291,85)[]{$\tilde{\tau}^-_i$}
\DashLine(280,25)(255,50){4}
\Text(291,15)[]{$\tilde{\tau}^+_j$}

\end{picture}\\
\end{center}
\smallskip
\noindent
{\bf Figure~1}: {\it The mechanisms contributing to the process
            $\mu^+\mu^-\rightarrow \tilde{\tau}^-_i\tilde{\tau}^+_j$:
	    three spin--0 neutral--Higgs boson exchanges
	    and spin--1 $\gamma$ and $Z$ exchanges. The 
	    indices $i,j$ are 1 or 2, and the index $k$
	    is 1,2 or 3. }
\bigskip 
\bigskip 

\subsection{Production amplitudes}
\label{subsec:amplitude}

As shown in Fig.~1, the matrix element ${\cal M}_{ij}$ for
$\mu^+\mu^-\rightarrow \tilde{\tau}_i^- \tilde{\tau}^+_j$ receives
contributions from $\gamma$ and $Z$ exchange as well as from the
exchange of the neutral Higgs bosons of the MSSM. It can be written
as:
\begin{eqnarray}
{\cal M}_{ij}=\frac{e^2}{s}\sum_{\alpha=\pm}\bigg\{
Z_{\alpha ij}\left[\,\bar{v}(\bar{p})(\not\!{p}_i-\not\!{p}_j)
                   \,P_\alpha u(p)\right]
 +m_W H_{\alpha ij}\left[\,\bar{v}(\bar{p})\, P_\alpha u(p)
                   \right]\bigg\}\,,
\label{eq:production amplitude}
\end{eqnarray}
where $\alpha=\pm$, $P_\pm=(1\pm \gamma_5)/2$, and the vector and
scalar chiral couplings are given by
\begin{eqnarray} \label{eq:couplings}
&& Z_{-ij}=\delta_{ij}
       +D_Z(s)\frac{s^2_W-1/2}{c^2_W s^2_W}\,Q^Z_{ij}\,,\nonumber\\
&& Z_{+ij}=\delta_{ij}
       +D_Z(s)\frac{1}{c^2_W}\,Q^Z_{ij}\,,\nonumber\\
&& H_{-ij}=-Y_\mu\,\frac{D_{H_k}(s)}{c^2_W s^2_W}\,V_{k;ij}
          \left[O_{2k}+i s_\beta O_{1k}\right]
          \,,\nonumber\\[1mm]
&& H_{+ij}=-Y_\mu\,\frac{D_{H_k}(s)}{c^2_W s^2_W}\,V_{k;ij}
          \left[O_{2k}-i s_\beta O_{1k}\right]\,.
\end{eqnarray}
The coupling $Y_\mu$, and the $Z$ and $H_k$ propagators $D_Z$ and
$D_{H_k}$ ($k=1,2,3$) are:
\begin{eqnarray}
Y_\mu       &=& \frac{m_\mu}{2m_W\, c_\beta}\,,\nonumber\\ 
D_Z(s)      &=& \frac{s}{s-m^2_Z+im_Z\Gamma_Z}\,,\nonumber\\
D_{H_k}(s)  &=& \frac{s}{s-m^2_{H_k}+im_{H_k}\Gamma_{H_k}}\,.
\end{eqnarray}

Defining the polar angle of the flight direction of the tau slepton
$\tilde{\tau}^-_i$ with respect to the $\mu^-$ beam direction by
$\Theta$ (see Fig.~2), the explicit form of the production amplitude
(\ref{eq:production amplitude}) can be evaluated in the helicity basis
by the 2--component spinor technique of Ref.~\cite{HZ}\footnote{Our
convention for the momentum--dependent Weyl spinor for fermions going
in $-z$ direction differs by an overall sign from that of
Ref.~\cite{HZ}}. We denote the $\mu^-$ and $\mu^+$ helicities by
$\sigma$ and $\bar\sigma$, with $\sigma = +$ and $-$ standing for
right-- and left--handed particles, respectively. Neglecting the muon
mass in the kinematics, the helicity amplitudes
\begin{eqnarray}
{\cal M}_{ij}(\sigma, \bar{\sigma}) \equiv 
 e^2\,\langle\,\sigma\bar{\sigma}\rangle_{ij}, 
\end{eqnarray}
read
\begin{eqnarray}
&& \langle ++\rangle_{ij}  =-\frac{m_W}{\sqrt{s}}\, H_{+ij}\,,\qquad
   \langle +-\rangle_{ij}  =-Z_{+ij}\,\beta\,\sin\Theta\,,\nonumber\\
&& \langle --\rangle_{ij}  =+\frac{m_W}{\sqrt{s}}\, H_{-ij}\,,\qquad
   \langle -+\rangle_{ij}  =-Z_{-ij}\,\beta\,\sin\Theta\, .
\label{eq:helicity amplitude}
\end{eqnarray}
Here $\beta=\lambda^{1/2} (1, m^2_{\tilde{\tau}^-_i}/s,
m^2_{\tilde{\tau}^+_j}/s)$, with $\lambda(x,y,z) = x^2 + y^2 + z^2 -
2(xy+yz+zx)$, describes the 3--momentum of the scalar $\tau$ leptons
in the center--of--mass frame. Note that in the limit of negligible
kinematic mass of the muons vector boson exchange only contributes to
configurations where $\mu^-$ and $\mu^+$ have opposite helicities,
whereas Higgs boson exchange only contributes to configurations with
equal $\mu^+$ and $\mu^-$ helicities. This implies that Higgs and
gauge boson exchange contributions can only interfere with each other
if at least one of the muons has nonvanishing {\em transverse}
polarization, which can be understood as {\em coherent} superposition
of left-- and right--handed helicity states \cite{HZ}; see
Eqs.~(\ref{eq:trace}) and (\ref{eq:rho}) below. 

\bigskip

\vspace*{0cm} 
\begin{center}
\begin{picture}(300,100)(0,0)

\Line(15,10)(55,90)
\Line(55,90)(285,90)
\Line(285,90)(245,10)
\Line(245,10)(15,10)

\Text(-3,50)[]{$\mu^-$}
\Line(15,50)(145,50)
\Line(145,50)(140,54)
\Line(145,50)(140,47)

\Text(298,50)[]{$\mu^+$}
\Line(155,50)(285,50)
\Line(155,50)(160,54)
\Line(155,50)(160,47)

\Vertex(150,50){3}

\Text(209,109)[]{$\tilde{\tau}^-_i$}
\Line(150,50)(200,100)
\Line(200,100)(200,95)
\Line(200,100)(197,100)

\Text(93,-7)[]{$\tilde{\tau}^+_j$}
\Line(150,50)(100,0)
\Line(100,0)(105,0)
\Line(100,0)(100,5)

\Text(90,65)[]{$\alpha$}
\ArrowArc(75,50)(20,60,90)
\Text(232,80)[]{$\bar{\alpha}$}
\ArrowArc(235,50)(20,60,135)

\Text(62,79)[]{$P_{_T}$}
\DashLine(75,50)(95,90){3}

\Text(195,79)[]{$\bar{P}_{_T}$}
\DashLine(235,50)(255,90){3}

\Text(190,62)[]{$\Theta$}
\ArrowArc(150,50)(30,0,45)

\SetWidth{2}

\Line(75,50)(75,80)
\Line(75,80)(71,77)
\Line(75,80)(79,77)

\Line(235,50)(205,80)
\Line(205,80)(210,80)
\Line(205,80)(205,75)

\end{picture}\\
\end{center}
\bigskip
\noindent
{\bf Figure~2}: {\it Schematic depiction of the production plane
            with the polar scattering angle $\Theta$.
	    The transverse polarization vectors $P_T$ and $\bar{P}_T$
	    have azimuthal angles $\alpha$ and $\bar{\alpha}$
	    with respect to the event plane, respectively.}
\bigskip
\bigskip

Before turning to the calculation of the cross section, let us briefly
describe the symmetry properties of the helicity amplitudes.
The CP transformation leads to a relation among the transition
helicity amplitudes:
\begin{eqnarray}
\langle \sigma\bar{\sigma}\rangle_{ij} 
  &\stackrel{\rm CP}{\longleftrightarrow} &
  (-1)(-1)^{(\sigma-\bar{\sigma})/2}
  \,\langle -\bar{\sigma},-\sigma\rangle_{ji} \, ;
\end{eqnarray}
note that $\sigma - \bar\sigma$ gives the total spin, if the $\mu^-$
momentum direction is used as quantization axis. Equivalently one has
for the scalar and vector helicity amplitudes:
\begin{eqnarray} \label{eq:cptrafo2}
\langle \pm\pm\rangle_{ij}\ \  
  \stackrel{\rm CP}{\longleftrightarrow} \ \ 
   -\langle \mp\mp\rangle_{ji}\,,\qquad
\langle \pm\mp\rangle_{ij} \ \ 
  \stackrel{\rm CP}{\longleftrightarrow} \ \ 
   +\langle \pm\mp\rangle_{ji}\,.
\end{eqnarray}
Eqs.~(\ref{eq:cptrafo2}), (\ref{eq:couplings}) and (\ref{eq:helicity
amplitude}) show that CP violation in diagonal channels ($i=j$) can
only occur in the presence of nonvanishing pseudoscalar couplings,
since the contributions $\propto O_{1k}$ in Eq.~(\ref{eq:couplings})
are the only ones that violate relation (\ref{eq:cptrafo2}). We will
see in the next section that observable CP--odd effects can only arise
in diagonal $\tilde\tau$ pair production if a pseudoscalar coupling
interferes with one of the other couplings. On the other hand, all
four sets of couplings (scalar, pseudoscalar, vector, and axial
vector) can give rise to CP violation in non--diagonal $\tilde\tau$
pair production ($i \neq j$). 

Another useful classification is provided by the 
so--called ``naive" time reversal $\tilde{\rm T}$. Like the proper
T-transformation, this transformation changes the directions of all
3--momenta and spin vectors, i.e. it leaves helicities
unchanged. However, unlike a T transformation, a $\tilde{\rm T}$
transformation does {\em not} interchange initial and final states. In
our case a $\tilde{\rm T}$ transformation simply corresponds to taking
the complex conjugate of the helicity amplitudes. Under the 
combined CP$\tilde{\rm T}$ transformation
the helicity amplitudes therefore transform as follows:
\begin{eqnarray}
\langle \pm\pm\rangle_{ij} \ \ 
  \stackrel{\rm CP\tilde{\rm T}}{\longleftrightarrow} \ \ 
   -\langle \mp\mp\rangle^*_{ji}\,,\qquad
\langle \pm\mp\rangle_{ij} \ \ 
  \stackrel{\rm CP\tilde{\rm T}}{\longleftrightarrow} \ \ 
   +\langle \pm\mp\rangle^*_{ji}\,.
\end{eqnarray}
We note that tree--level CP$\tilde{\rm T}$ violation is only possible
in the presence of finite $Z$ or Higgs boson widths. It is therefore
very useful to analyze the CP and CP$\tilde{\rm T}$ properties of any
physical observable simultaneously, so as to investigate not only CP
violation itself but also its dependence on the real or imaginary
part of some product(s) of propagators. 

\subsection{Polarized production cross sections} 

The production cross section for general (longitudinal or transverse)
beam polarization can be computed either using standard trace
techniques (employing general spin projection operators), or from the
helicity amplitudes by a suitable rotation \cite{HZ} from the helicity
basis to a general spin basis. In the former case, neglecting the muon
mass in the spin projection operators we can obtain the following
approximated form for the $\mu^\mp$ projection operators:
\begin{eqnarray}
\frac{1}{2}(\not\!{p}+m)(1+\gamma_5\not\!{s})\longrightarrow
 \frac{1}{2}(1+P_L\gamma_5)\not\!{p}+\frac{1}{2}\gamma_5P_T(\cos\alpha
      \not\!{n}_1+\sin\alpha\not\!{n}_2)\not\!{p}\,,\nonumber\\
\frac{1}{2}(\not\!{\bar{p}}-m)(1+\gamma_5\not\!{\bar{s}})
   \longrightarrow
 \frac{1}{2}(1-\bar{P}_L\gamma_5)\not\!{\bar{p}}
 +\frac{1}{2}\gamma_5\bar{P}_T(\cos\bar{\alpha}\not\!{n}_1
  +\sin\bar{\alpha}\not\!{n}_2)\not\!{\bar{p}}\,.
\label{eq:projection}
\end{eqnarray}
Here $s_\mu$ is the spin 4--vector, and $n_{1\mu}$ and $n_{2\mu}$ are
purely spatial vectors satisfying $n_i \cdot n_j = -\delta_{ij}$ and
$n_i \cdot p = n_i \cdot \bar p = 0$. $P_L$ and $\bar{P}_L$ are the
longitudinal polarizations of the $\mu^-$ and $\mu^+$ beams, while
$P_T$ and $\bar{P}_T$ are the degrees of transverse polarization with
$\alpha$ and $\bar{\alpha}$ being the azimuthal angles between the
transverse polarization vectors and the momentum vector of
$\tilde{\tau}^-_i$ as shown in Fig.~2.  We also note that
$P^2_L+P^2_T\,\leq\, 1$ and $\bar{P}^2_L +\bar{P}^2_T\,\leq\, 1$.

Equivalently the polarization weighted matrix element squared in the
helicity basis is given by
\begin{eqnarray}
\overline{\sum}|{\cal M}|^2
 =\sum_{\sigma\sigma'\bar{\sigma}\bar{\sigma}'}
  {\cal M}_{\sigma\bar{\sigma}} {\cal M}^*_{\sigma'\bar{\sigma}'}
  \rho^-_{\sigma\sigma'}\rho^+_{\bar{\sigma}\bar{\sigma}'}
 ={\rm Tr}\left[{\cal M}\,\rho^+{\cal M}^\dagger\rho^{-T}\right]\,,
 \label{eq:trace}
\end{eqnarray}
where ${\cal M}_{\sigma\bar{\sigma}}$ ($\sigma,\bar{\sigma}=\pm$)
denotes the helicity amplitude for any given production process 
$\mu^-(\sigma)\mu^+(\bar{\sigma})\rightarrow X$ and
the $2\times 2$ matrices $\rho^\mp$ are the polarization
density matrices for the initial $\mu^\mp$ beams:
\begin{eqnarray} \label{eq:rho}
\rho^-=\frac{1}{2}\left(\begin{array}{cc}
     1+P_L                  & P_T\,{\rm e}^{-i\alpha}  \\[2mm]
    P_T\,{\rm e}^{i\alpha} & 1-P_L
                   \end{array}\right)\,,\qquad
\rho^+=\frac{1}{2}\left(\begin{array}{cc}
     1+\bar{P}_L       & -\bar{P}_T\,{\rm e}^{i\bar{\alpha}}  \\[2mm]
    -\bar{P}_T\,{\rm e}^{-i\bar{\alpha}} & 1-\bar{P}_L
                   \end{array}\right)\,.
\end{eqnarray}

Applying the projection operators (\ref{eq:projection}) or/and 
evaluating the trace (\ref{eq:trace}) leads to the following matrix 
element squared for polarized $\mu^+ \mu^- \rightarrow \tilde\tau_i^-
\tilde\tau_j^+$ production:
\begin{eqnarray} \label{eq:Sigma}
 &&\Sigma_{ij} \equiv\sum_{\sigma\sigma'\bar{\sigma}\bar{\sigma}'}
     \langle\sigma\bar{\sigma}\rangle_{ij}\,
     \langle\sigma'\bar{\sigma'}\rangle^*_{ij}\,
     \rho^-_{\sigma\sigma'} \rho^+_{\bar{\sigma}\bar{\sigma}'}
     \nonumber \\
 &&\hskip 0.7cm = \left(1-P_L\bar{P}_L\right)\, C_1[ij]
   + \left( P_L-\bar{P}_L\right)\, C_2[ij] \nonumber \\
 &&\hskip 0.8cm + \left(1+P_L\bar{P}_L\right)\, C_3[ij] 
   + \left( P_L+\bar{P}_L\right)\, C_4[ij]\nonumber \\
 &&\hskip 0.8cm +\,(P_T\cos\alpha+\bar{P}_T\cos\bar{\alpha})\, C_5[ij] 
                +\,(P_T\sin\alpha+\bar{P}_T\sin\bar{\alpha})\, C_6[ij] 
		\nonumber\\[1mm]
 &&\hskip 0.8cm +\,(P_T\cos\alpha-\bar{P}_T\cos\bar{\alpha})\, C_7[ij] 
                +\,(P_T\sin\alpha-\bar{P}_T\sin\bar{\alpha})\, C_8[ij] 
                \nonumber \\[1mm]
 &&\hskip 0.8cm 
 +\,(P_L\bar{P}_T\cos\bar{\alpha}+\bar{P}_L P_T\cos\alpha)\,\, C_9[ij]
\,+\,(P_L\bar{P}_T\sin\bar{\alpha}+\bar{P}_L P_T\sin\alpha)\,\, 
  C_{10}[ij] \nonumber\\[1mm]
 &&\hskip 0.8cm 
+\,(P_L\bar{P}_T\cos\bar{\alpha}-\bar{P}_L P_T\cos\alpha)\, C_{11}[ij]
+\,(P_L\bar{P}_T\sin\bar{\alpha}-\bar{P}_L P_T\sin\alpha)\, C_{12}[ij]
                 \nonumber \\[1mm]
&&\hskip 0.8cm 
+\,P_T\bar{P}_T\left[\,\cos(\alpha+\bar{\alpha})\, C_{13}[ij]
+ \sin(\alpha+\bar{\alpha})\, C_{14}[ij]\right]\nonumber\\[1mm]
&&\hskip 0.8cm 
+\,P_T\bar{P}_T\left[\,\cos(\alpha-\bar{\alpha})\, C_{15}[ij]
   + \sin(\alpha-\bar{\alpha})\, C_{16}[ij]\right] .
\end{eqnarray}
The coefficients $C_n$ ($n = 1$ - $16$) are defined in terms
of the helicity amplitudes by
{\small 
\begin{eqnarray*} \label{eq:cij}
C_1&=&\frac{1}{4}
          \left[|\langle +-\rangle|^2 
              + |\langle -+\rangle|^2\right], \ \
   C_2\, =\, \frac{1}{4}
          \left[|\langle +-\rangle|^2
              - |\langle -+\rangle|^2\right], \nonumber\\
C_3&=&\frac{1}{4}
          \left[|\langle ++\rangle|^2
              + |\langle --\rangle|^2\right], \ \
   C_4\, =\, \frac{1}{4}
          \left[|\langle ++\rangle|^2
              - |\langle --\rangle|^2\right], \nonumber\\
C_5&=&\frac{1}{4}\,\real\left(\langle ++\rangle
    - \langle --\rangle\right)
    \left(\langle -+\rangle-\langle +-\rangle\right)^*, \nonumber\\ 
C_6&=&\frac{1}{4}\,\imag \left(\langle ++\rangle
    - \langle --\rangle\right)
    \left(\langle -+\rangle+\langle +-\rangle\right)^*, \nonumber\\ 
C_7&=&\frac{1}{4}\,\real \left(\langle ++\rangle
    + \langle --\rangle\right)
    \left(\langle -+\rangle + \langle +- \rangle\right)^*, \nonumber\\ 
C_8&=&\frac{1}{4}\,\imag \left(\langle ++\rangle
    + \langle --\rangle\right)
    \left(\langle -+\rangle -\langle +-\rangle\right)^*,\nonumber\\ 
C_9&=&\frac{1}{4}\,\real \left(\langle ++\rangle
    + \langle --\rangle\right)
    \left(\langle -+\rangle-\langle +-\rangle\right)^*, \nonumber\\ 
C_{10}&=&\frac{1}{4}\,\imag
\left(\langle --\rangle + \langle ++\rangle\right)
\left(\langle -+\rangle + \langle +-\rangle\right)^*,   \nonumber\\ 
C_{11}&=&\frac{1}{4}\,\real
\left(\langle --\rangle
    - \langle ++\rangle\right)
\left(\langle -+\rangle
    + \langle +-\rangle\right)^*,   \nonumber\\ 
C_{12}&=&\frac{1}{4}\,\imag
\left(\langle ++\rangle
    - \langle --\rangle\right)
\left(\langle +-\rangle
    - \langle -+\rangle\right)^*,   \nonumber\\ 
C_{13}&=&-\frac{1}{2}\,\real
\left[\langle -+\rangle
      \langle +-\rangle^*\right],\ \
   C_{14}\, =\, \frac{1}{2}\,\imag
\left[\langle -+\rangle
      \langle +-\rangle^*\right],   \nonumber\\
C_{15}&=&-\frac{1}{2}\,\real
\left[\langle --\rangle
      \langle ++\rangle^*\right],\ \      
    C_{16}\, =\, \frac{1}{2}\,\imag
\left[\langle --\rangle
      \langle ++\rangle^*\right].      
\end{eqnarray*}
}
{ }\\[-2.33cm]
\begin{eqnarray}
{ } 
\end{eqnarray}\\
The production cross section is then given in terms of the
distribution $\Sigma_{ij}$ in Eq.~(\ref{eq:Sigma}) by
\begin{eqnarray}
\frac{d\sigma}{d\cos\Theta\, d\Phi}
\left(\mu^+\mu^-\rightarrow\tilde{\tau}^-_i \tilde{\tau}^+_j\right)
  =\frac{\alpha^2}{4\,s}\,\beta\,\Sigma_{ij}\,,
\end{eqnarray}
with $\beta$ as in Eqs.~(\ref{eq:helicity amplitude}). The dependence
of the distribution $\Sigma_{ij}$ on the azimuthal angle $\Phi$ of the
production plane is encoded in the angles $\alpha$ and
$\bar{\alpha}$. If the azimuthal angle $\Phi$ is measured with respect
to the direction of the $\mu^-$ transverse polarization vector, the
$\Phi$ dependence can be exhibited explicitly by taking for the angles
$\alpha$ and $\bar{\alpha}$
\begin{eqnarray} \label{eq:alpha}
\alpha=-\Phi,\qquad 
\bar{\alpha}=\eta -\Phi\,,
\end{eqnarray}
where $\eta$ is the {\it rotational invariant} difference
$\bar{\alpha}-\alpha$ of the azimuthal angles of the $\mu^+$ and
$\mu^-$ transverse polarization vectors. The polarization
coefficients can be classified according to their correlation patterns
of the vector and scalar contributions as follows:
\begin{eqnarray}
\begin{array}{lcl}
{\rm Vector\ \ and\ \ vector\, (VV)\ \ correlations}      & : & 
      C_1,\,C_2,\,C_{13},\,C_{14},\\[1mm]
{\rm Scalar\ \ \, and\ \ scalar\, (\,SS\,\,) \ \ correlations} & : & 
      C_3,\,C_4,\,C_{15},\,C_{16},\\[1mm]
{\rm Scalar\ \ \, and\ \ vector\, (\,SV)\ \ correlations}      & : &
      C_5,\,C_6,\,C_7,\,C_8,\,C_9,\,C_{10},\,C_{11},\,C_{12}.
\end{array}
\end{eqnarray}
Among the sixteen observables, only six observables - the 2 VV
observables $\{C_1,C_2\}$ and the 4 SS observables $\{C_3,C_4, C_{15},
C_{16}\}$ - can be measured independently of the azimuthal angle
$\Phi$, but the other 10 observables require the reconstruction of
the production plane. This is true in particular for the observables
involving scalar--vector correlations; as already stated in Sec.~3.2,
these observables are only nonzero if at least one of the incoming
beams has nonvanishing transverse polarization. 

\subsection{Observables with definite CP and CP$\tilde{\rm T}$ 
            properties}

The CP and CP$\tilde{\rm T}$ transformations act on the polarization 
vectors of the initial muon beams according to the simultaneous 
exchanges:
\begin{eqnarray}
\label{eq:transformation}
&&  P_L \ \
   \stackrel{\rm CP}{\longleftrightarrow} \ \ 
   -\bar{P}_L,\quad
    P_T \ \
   \stackrel{\rm CP}{\longleftrightarrow} \ \ 
    \bar{P}_T\, ;\qquad 
    \alpha\ \ 
   \stackrel{\rm CP}{\longleftrightarrow} \ \ 
    \,\bar{\alpha}\,,\nonumber\\[2mm]
&&  P_L \ \
   \stackrel{\rm CP\tilde{\rm T}}{\longleftrightarrow} \ \ 
   -\bar{P}_L,\quad
    P_T \ \
   \stackrel{\rm CP\tilde{\rm T}}{\longleftrightarrow} \ \ 
    \bar{P}_T\, ;\qquad 
    \alpha \ \
   \stackrel{\rm CP\tilde{\rm T}}{\longleftrightarrow}  
    -\bar{\alpha}\,.
\end{eqnarray}
The CP and CP$\tilde{\rm T}$ relations (\ref{eq:transformation}) of
the polarization vectors lead to the following classification of the
terms in Eq.~(\ref{eq:Sigma}): 7 CP--even and CP$\tilde{\rm T}$--even,
3 CP--even and CP$\tilde{\rm T}$--odd, 3 CP--odd and CP$\tilde{\rm
T}$--even, and 3 CP--odd and CP$\tilde{\rm T}$--odd polarization
factors, as shown in Table~\ref{tab:polarization factors}.

\begin{table}[\hbt]
\caption{\label{tab:polarization factors}
{\it The classification of the correlations of the muon and anti--muon
     polarization vectors that appear in Eq.~(\ref{eq:Sigma}) according
     to their CP and CP$\tilde{T}$ 
     properties, based on the CP and CP$\tilde{T}$ relations
     (\ref{eq:transformation}).}}
\mbox{ }\hskip 0.2cm
\begin{center}
\begin{tabular}{|c|c|l|}\hline
  CP & CP$\tilde{\rm T}$ & 
       { }\hskip 2cm Polarization factors\\\hline \hline
 &  &  \\[-3mm]
 even    &  even     & $1-P_L\bar{P}_L$,\ \ \  $P_L-\bar{P}_L$, \ \ 
                       $1+P_L\bar{P}_L$, \ \ 
		       $P_T\cos\alpha+\bar{P}_T\cos
		        \bar{\alpha}$ \\[2mm]
         &           & $P_L\bar{P}_T\cos\bar{\alpha} 
	                - \bar{P}_L P_T\cos\alpha$,
                       $P_T\bar{P}_T\cos(\alpha+\bar{\alpha})$, 
                       $P_T\bar{P}_T\cos(\alpha-\bar{\alpha})$ \\
 & & \\[-3mm]
\cline{2-3} 
 & & \\[-3mm]
         &  odd      & $P_T\sin\alpha+\bar{P}_T\sin\bar{\alpha}$,
	               $P_L\bar{P}_T\sin\bar{\alpha} 
	                - \bar{P}_L P_T\sin\alpha$,
                    $P_T\bar{P}_T\sin(\alpha+\bar{\alpha})$ \\[0.5mm] 
		       \hline \hline
 & & \\[-3mm]
 odd     &  even     & $P_T\sin\alpha-\bar{P}_T\sin\bar{\alpha}$,
	               $P_L\bar{P}_T\sin\bar{\alpha} 
	                + \bar{P}_L P_T\sin\alpha$,
                    $P_T\bar{P}_T\sin(\alpha-\bar{\alpha})$ \\[1mm] 
 & & \\[-4mm]
\cline{2-3} 
 & & \\[-3mm]
         & odd       & $P_L+\bar{P}_L$,\ \ 
	               $P_T\cos\alpha-\bar{P}_T\cos\bar{\alpha}$,\ \
	               $P_L\bar{P}_T\cos\bar{\alpha} 
	                + \bar{P}_L P_T\cos\alpha$ \\[2mm] 
\hline
\end{tabular}
\end{center}
\end{table}

The coefficients $C_n[ij]$ ($n=1$-16) corresponding to these
polarization factors are bilinears in $Z_{\pm ij}$ and $H_{\pm ij}$.
For later convenience we define the asymmetric combinations
$[\,C_n[12]]$ and the symmetric combinations $\{C_n[ij]\}$
$(ij=11,12,22$) as
\begin{eqnarray}
[\,C_n[12]]\equiv\frac{1}{2}\left(\,C_n[12] - C_n[21]\,\right),\qquad 
\{C_n[ij]\}\equiv\frac{1}{2}\left(\,C_n[ij] + C_n[ji]\,\right).
\end{eqnarray}
Clearly the coefficients multiplying CP--even polarization factors
(the first group in Table~1) can give rise to CP violation only
through the antisymmetric combinations $[C_n[12]]$, which describe
rate asymmetries. In contrast, all symmetric combinations
$\{C_n[ij]\}$ of the coefficients of the second group in Table 1,
which have CP--odd polarization factors, can contribute to CP--odd
polarization or azimuthal angle asymmetries; these can be probed for
three different CP--even final states ($\tilde{\tau}_i^-
\tilde{\tau}_i^+,\ i=1,2$ and the sum of $\tilde{\tau}_1^-
\tilde{\tau}_2^+$ and $\tilde{\tau}_1^+ \tilde{\tau}_2^-$ production).
Based on these observations one can classify all the 64 distributions
according to their CP and CP$\tilde{\rm T}$ parities as shown in
Table~\ref{tab:distribution}. We will show in Sec.~4.2 that all 64
coefficients appearing in Eq.~(\ref{eq:Sigma}) can be extracted
independently, {\em if} the polarization of both beams can be be
controlled completely and the final tau slepton pair is identified. 

\begin{table}[\hbt]
\caption{\label{tab:distribution}
{\it The classification of the coefficients $C_n[ij]$ ($n=1$-$16$,
      $ij=11,12,21,22$) according to their 
      CP and CP$\tilde{\rm T}$ properties. The underlined observables
      can be constructed without direct reconstruction of the 
      scattering plane. Here, $\{C_n[ij]\}=(C_n[ij]+C_n[ji])/2$ and 
      $[\,C_n[12]]=(C_n[12]-C_n[21])/2$.}}
\begin{center}
\begin{tabular}{|c|c|l|}\hline
  CP & CP$\tilde{\rm T}$ & 
       { }\hskip 2cm Coefficients $(ij =11,12,22)$\\[-0.7mm]
       \hline \hline
 &  &  \\[-3mm]
 even    &  even   & $\underline{\{C_1[ij]\}}$, 
                     $\underline{\{C_2[ij]\}}$,
                     $\underline{\{C_3[ij]\}}$, 
		     $\{C_5[ij]\}$, 
                     $\{C_{11}[ij]\}$, $\{C_{13}[ij]\}$\\[2mm]
         &         & $\underline{\{C_{15}[ij]\}}$, 
	             $\underline{[\,C_4[12]]}$, 
		     $[\,C_7[12]]$, $[\,C_9[12]\,]$ \\
 & & \\[-2mm]
\cline{2-3} 
 & & \\[-2mm]
         &  odd      & 
	               $\{C_6[ij]\}$, $\{C_{12}[ij]\}$, 
	               $\{C_{14}[ij]\}$, 
                       $[\,C_8[12]]$, $[\,C_{10}[12]]$, 
	               $\underline{[\,C_{16}[12]]}$\\[1mm] 
		       \hline \hline
 & & \\[-3mm]
 odd     &  even     & $\{C_8[ij]\}$, $\{C_{10}[ij]\}$, 
	               $\underline{\{C_{16}[ij]\}}$, 
                       $[\,C_6[12]]$, $[\,C_{12}[12]]$, 
	               $[\,C_{14}[12]]$\\[0.5mm] 
 & & \\[-3mm]
\cline{2-3} 
 & & \\[-3mm]
         & odd       & $\underline{\{C_4[ij]\}}$, 
	               $\{C_7[ij]\}$, 
	               $\{C_9[ij]\}$, 
		       $\underline{[\,C_1[12]]}$, 
		       $\underline{[\,C_2[12]]}$, 
		       $\underline{[\,C_3[12]]}$\\[2mm]
         &           & $[\,C_5[12]]$, $[\,C_{11}[12]]$, 
	               $[\,C_{13}[12]]$, 
		       $\underline{[\,C_{15}[12]]}$\\[2mm] 
\hline
\end{tabular}
\end{center}
\end{table}

We thus see that one can define 10 CP--violating rate asymmetries
involving anti--symmetric combinations in the $\tilde{\tau}$ indices
and 6 polarization/angle asymmetries involving symmetric combinations
in the $\tilde{\tau}$ indices; we recall that the latter can be
studied for three different final states, leading to a total of 28
different asymmetries.  More detailed explanations of the definition
of the rate and polarization asymmetries and of the extraction of the
asymmetries by adjusting the muon and anti--muon polarizations will be
given in the next section.

Any CP$\tilde{\rm T}$--odd observable requires some CP--preserving
phase. At tree--level this can only be provided by the $Z$ and Higgs
boson widths in the tau slepton pair production processes. Therefore,
if the $Z$ width is neglected, in which case $Z_{\pm ij}$ is
Hermitian, the following six quantities involving only the $\gamma$ or
$Z$ boson couplings to muons and tau sleptons vanish\,:
\begin{eqnarray} \label{eq:zero1}
[\,C_1[12]\,]=[\,C_2[12]\,]=[\,C_{13}[12]\,]=C_{14}[11] = C_{14}[22] =
0.
\end{eqnarray}
Furthermore, Eqs.~(\ref{eq:qz}), (\ref{eq:couplings}) and (\ref{eq:cij})
show that  
\begin{eqnarray} \label{eq:zero2}
[\,C_{14}[12]\,] = \{C_{14}[12]\} = 0,
\end{eqnarray}
even for nonvanishing $\Gamma_Z$.
All other CP$\tilde{\rm T}$--odd quantities can get contributions from
some Higgs boson width, as will be clarified in the next section. 

\subsection{CP$\tilde{\rm T}$--even and odd combinations of 
            Higgs boson propagators}

The behavior of each observable under a CP$\tilde{\rm T}$
transformation plays a crucial role in determining the interference
pattern of the Higgs boson contributions, among themselves and with
the $\gamma$ and $Z$ boson contributions to this observable. Every
CP$\tilde{\rm T}$--even (odd) observable involving only scalar
contributions depends on the real (imaginary) parts of the products
$D_{H_k}D^*_{H_l}$ of Higgs boson propagators; note that only $k \neq
l$ gives a nonzero imaginary part. We already showed that the
observable $C_{14}$ involving only the vector contributions is
negligible, see Eqs.~(\ref{eq:zero1}) and (\ref{eq:zero2}). In the
approximation $\Gamma_Z = 0$ all the vector boson contributions $Z_-$
and $Z_+$ in Eq.~(\ref{eq:couplings}) are hermitian. As a result, the
interference terms between scalar and vector contributions depend on
the real or imaginary parts of the Higgs boson propagators $D_{H_k}$
themselves.

This discussion shows that it is convenient to introduce abbreviated
notations for the real and imaginary parts:
\begin{eqnarray} \label{eq:ssprops}
&& {\cal S}_{kl}\equiv\,\, \real [D_{H_k} D^*_{H_l}]
                \, = s^2\frac{(s-m^2_{H_k})(s-m^2_{H_l})
		       +m_{H_k}m_{H_l}\Gamma_{H_k}\Gamma_{H_l}}{
		        [(s-m^2_{H_k})^2+m^2_{H_k}\Gamma^2_{H_k}]
		        [(s-m^2_{H_l})^2+m^2_{H_l}\Gamma^2_{H_l}]},
			\nonumber\\
&& {\cal D}_{kl}\equiv\imag [D_{H_k} D^*_{H_l}]
                = s^2 \frac{(s-m^2_{H_k})\, m_{H_l}\Gamma_{H_l}
		       -(s-m^2_{H_l})\, m_{H_k}\Gamma_{H_k}}{
		        [(s-m^2_{H_k})^2+m^2_{H_k}\Gamma^2_{H_k}]
		        [(s-m^2_{H_l})^2+m^2_{H_l}\Gamma^2_{H_l}]}.
\end{eqnarray}
It is worthwhile to note two points: (a) the denominator reveals a
typical two--pole structure so that the Higgs boson contributions are
greatly enhanced near the poles; (b) the numerator of ${\cal S}_{kl}$
is negative between two resonances with a mass splitting larger than
their typical widths but positive otherwise. In contrast, for
realistic widths of the MSSM Higgs bosons the numerator of ${\cal
D}_{kl}$ is always positive (negative) if $m_{H_l} \geq m_{H_k}$
($m_{H_k}\geq m_{H_l}$). Moreover, it increases (decreases) linearly
with $s$ if $m_{H_l}\Gamma_{H_l} > \ (<) m_{H_k}\Gamma_{H_k}$.

As noted earlier, in the approximation $\Gamma_Z=0$ the SV
interference terms depend on the real and imaginary parts of each
Higgs boson propagators $D_{H_k}$, which are given by
\begin{eqnarray}
\real[D_{H_k}]=\frac{s\,(s-m^2_{H_K})}{(s-m^2_{H_k})^2
              + m^2_{H_k}\Gamma^2_{H_k}}, \qquad
\imag[D_{H_k}]=-\frac{s\, m_{H_K}\Gamma_{H_k}}{(s-m^2_{H_k})^2
              + m^2_{H_k}\Gamma^2_{H_k}},
\end{eqnarray}
respectively. Note that the real parts change their sign whenever 
the c.m. energy crosses the corresponding pole, but the imaginary 
parts are always negative.

Combining the coefficients from the mixing matrix elements with the
corresponding propagator--dependent factors enables us to
straightforwardly understand the qualitative $\sqrt{s}$ dependence of
the combinations of coefficients $C_n[ij]$ listed in Table~2. However,
the energy dependence of the corresponding properly normalized
asymmetries is also affected by the energy dependence of the
normalization factors, which can be nontrivial. We will discuss these
issues further in Sec.~5, where we present numerical results.

\section{Polarization and rate asymmetries}
\label{sec:reconstruction}
\setcounter{footnote}{1}

In this section, we present the explicit forms of all the VV, SS, 
and SV polarization and rate correlations and then explain how to 
extract the corresponding asymmetries by adjusting initial beam
polarization and, in some cases, inserting weight functions of the
polar angle in the phase space integral. For completeness we include
the CP--even correlations listed in the upper half of Table~2. 

\subsection{Expressions for the correlations}

\subsubsection{Vector--vector correlations}

The VV correlations are very useful for reconstructing the tau slepton
system independent of the Higgs boson system.  In principle there
exist 9 non--trivial CP--even and CP$\tilde{\rm T}$--even VV
observables $\left(\{C_{\,1}[ij]\}, \{C_{\,2}[ij]\},
\{C_{13}[ij]\}\right)$ and 1 CP--odd and CP$\tilde{\rm T}$--even
observable $[C_{14}[12]]$. However, we already saw that this latter
observable vanishes. Since the CP--violating phase $\phi_\tau$ in the
tau slepton system affects neither the masses of the $\tilde\tau$
eigenstates nor the absolute values of their gauge couplings, it
cannot be directly measured through VV correlations only.\footnote{The
phase $\Phi_{A_\tau}$ does affect the $\tilde \tau$ masses. However,
it could be extracted from measurements of $\tilde \tau$ masses and
$\theta_\tau$ only if $|A_\tau|$ and $|\mu| \tan\beta$ are already
known. In case of $|A_\tau|$ at least this seems quite unlikely.}
However, the absolute values of $X_\tau, Y_\tau$ and $Z_\tau$ defined
in Eq.~(\ref{eq:xyztau}) can be determined from the remaining 9
CP--even and CP$\tilde{\rm T}$--even observables, together with the
measurement of the two tau slepton masses. As a result, if $\tan\beta$
is determined (or constrained to be large) from other processes (e.g.,
chargino pair production in $e^+e^-$ collisions \cite{chipair}), the
above determination can be used to obtain the left/right
soft--breaking tau slepton masses $m_{\tilde{\tau}_{L,R}}$. The
absolute magnitude of $|A_\tau^* + \mu\, {\rm e}^{i\xi}\,\tan\beta|$
can be determined even if $\tan\beta$ is unknown. However, it is
necessary\footnote{The absolute value of $Z_\tau$ does depend on the
relative phase between $A_\tau^*$ and $\mu {\rm e}^{i\xi}$. However,
observables that only depend on $|Z_\tau|$ clearly cannot determine
this phase, unless $|A_\tau|$ and $|\mu \tan \beta|$ are already
known.} to consider different observables to determine the phase angle
$\phi_\tau={\rm arg}(-A_\tau-\mu^*\, {\rm e}^{-i\xi})$.

\subsubsection{Scalar--scalar correlations}

The scalar--scalar correlations involve not only tau slepton mixing but also
neutral Higgs boson mixing. However, it is known that the Higgs boson
mixing angles can be completely determined through on--shell
production of each Higgs boson with polarized muons \cite{eri}.  In
this section we therefore check the possibility of reconstructing the
tau slepton system through these scalar--scalar correlations.  There
are 16 such correlations: 7 CP--even and CP$\tilde{\rm T}$--even, 1
CP--even and CP$\tilde{\rm T}$--odd, 3 CP--odd and CP$\tilde{\rm
T}$--even and 5 CP--odd and CP$\tilde{\rm T}$--odd observables. It is
straightforward to obtain the explicit forms of the these 16
correlations:
\begin{eqnarray} \label{eq:sscorr}
&& \{C_3[ij]\} = +\left(\frac{Y_\mu}{c^2_W s^2_W}\right)^2
               \frac{m^2_W}{2s}\, {\cal S}_{kl}\,
	       \real \left(V_{k;ij}V^*_{l;ij}\right)
	       \left(O_{2k}O_{2l}+s^2_\beta\, O_{1k} O_{1l}\right),
	       \nonumber\\
&& [\, C_3[12]] =- \left(\frac{Y_\mu}{c^2_W s^2_W}\right)^2
               \frac{m^2_W}{2s}\, {\cal D}_{kl}\,
	       \imag \left(V_{k;12}V^*_{l;12}\right)
	       \left(O_{2k}O_{2l}+s^2_\beta\, O_{1k} O_{1l}\right),
	       \nonumber\\
&& \{C_4[ij]\} = +\left(\frac{Y_\mu}{c^2_W s^2_W}\right)^2
               \frac{m^2_W}{2s}\, {\cal D}_{kl}\,
	       \real \left(V_{k;ij}V^*_{l;ij}\right)
	       s_\beta \left(O_{1k}O_{2l}-O_{2k} O_{1l}\right),
	       \nonumber\\
&& [\, C_4[12]] =+ \left(\frac{Y_\mu}{c^2_W s^2_W}\right)^2
               \frac{m^2_W}{2s}\, {\cal S}_{kl}\,
	       \imag \left(V_{k;12}V^*_{l;12}\right)
	       s_\beta \left(O_{1k}O_{2l}-O_{2k} O_{1l}\right),
	       \nonumber\\
&& \{C_{15}[ij]\} = +\left(\frac{Y_\mu}{c^2_W s^2_W}\right)^2
               \frac{m^2_W}{2s}\, {\cal S}_{kl}\,
	       \real \left(V_{k;ij}V^*_{l;ij}\right)
	       \left(O_{2k}O_{2l}-s^2_\beta\, O_{1k} O_{1l}\right),
	       \nonumber\\
&& [\, C_{15}[12]] = -\left(\frac{Y_\mu}{c^2_W s^2_W}\right)^2
               \frac{m^2_W}{2s}\, {\cal D}_{kl}\,
	       \imag \left(V_{k;12}V^*_{l;12}\right)
	       \left(O_{2k}O_{2l}-s^2_\beta\, O_{1k} O_{1l}\right),
	       \nonumber\\
&& \{C_{16}[ij]\} = -\left(\frac{Y_\mu}{c^2_W s^2_W}\right)^2
               \frac{m^2_W}{2s}\, {\cal S}_{kl}\,
	       \real \left(V_{k;ij}V^*_{l;ij}\right)
	       s_\beta \left(O_{1k}O_{2l}+O_{2k} O_{1l}\right),
	       \nonumber\\
&& [\, C_{16}[12]] = +\left(\frac{Y_\mu}{c^2_W s^2_W}\right)^2
               \frac{m^2_W}{2s}\, {\cal D}_{kl}\,
	       \imag \left(V_{k;12}V^*_{l;12}\right)
	       s_\beta \left(O_{1k}O_{2l}+O_{2k} O_{1l}\right),
\label{eq:SS correlations}
\end{eqnarray}
where summation over the repeating indices $k$ and $l$ is understood.
In deriving Eqs.~(\ref{eq:sscorr}) we have exploited the relation
$V_{k;ij} = \left( V_{k;ji} \right)^*$. Note that any rate asymmetry,
defined to be proportional to the difference between the $[12]$ and
$[21]$ modes, is determined by the combination $\imag
(V_{k;12}V^*_{l;12})$, which is antisymmetric with respect to the
Higgs boson indices $k,l$. In addition, in the MSSM the lightest Higgs
boson has a mass of less than 130 GeV and a tiny width, so that its
contribution to the tau slepton pair production is negligible.  As a
result, only the propagator combinations ${\cal S}_{22}, \ {\cal
S}_{23}, \ {\cal S}_{33}$ and ${\cal D}_{23}$ are important in
determining the $\sqrt{s}$ dependence of these rate asymmetries. 

\subsubsection{Scalar--vector correlations}

All the SV correlations are proportional to the real or imaginary part
of a Higgs propagator, so that they also are sizable when the
c.m. energy is very close to one of the heavy Higgs boson
resonances.\footnote{$\real (D_{H_k}) = 0$ for $\sqrt{s} =
m_{H_k}$. However, since the mass splitting between the two heavy
Higgs states is small, the contribution from the second heavy Higgs
boson is usually still sizable at this point.}
\begin{eqnarray} \label{eq:svcorr}
&& \{C_5[ij]\} = +\frac{m_W Y_\mu \beta}{4\sqrt{s}c^4_W s^4_W}
     O_{2k} D_Z\, \real [D_{H_k}]\, 
     \real \left(V_{k;ij}\, Q^{Z*}_{ij}\right)\sin\Theta,
     \nonumber\\
&& [\, C_5[12]] = -\frac{m_W Y_\mu \beta}{4\sqrt{s}c^4_W s^4_W}
     O_{2k} D_Z\, \imag [D_{H_k}]\, 
     \imag \left(V_{k;12}\, Q^{Z*}_{12}\right)\sin\Theta,
     \nonumber\\
&& \{C_6[ij]\} = -\frac{m_W Y_\mu \beta}{\sqrt{s}c^2_W s^2_W}
     O_{2k} \imag [D_{H_k}]\,\real \left(V_{k;ij}\delta_{ij}
     +D_Z \frac{s^2_W-1/4}{c^2_W s^2_W} V_{k;ij}\,Q^{Z*}_{ij}\right)
     \sin\Theta,
     \nonumber\\
&& [\, C_6[12]] = -\frac{m_W Y_\mu \beta}{\sqrt{s}c^4_W s^4_W}
     (s^2_W-1/4) O_{2k} D_Z\, \real [D_{H_k}]\,
     \imag \left(V_{k;12}\,Q^{Z*}_{12}\right)
     \sin\Theta,
     \nonumber\\
&& \{C_7[ij]\} = -\frac{m_W Y_\mu \beta}{\sqrt{s}c^2_W s^2_W}
     s_\beta\, O_{1k} \imag [D_{H_k}]\,\real \left(V_{k;ij}\delta_{ij}
     +D_Z \frac{s^2_W-1/4}{c^2_W s^2_W} V_{k;ij}\,Q^{Z*}_{ij}\right)
     \sin\Theta,
     \nonumber\\
&& [\, C_7[12]] = -\frac{m_W Y_\mu \beta}{\sqrt{s}c^4_W s^4_W}
     (s^2_W-1/4) s_\beta O_{1k} D_Z\,\real [D_{H_k}]\,
     \imag \left(V_{k;12}\,Q^{Z*}_{12}\right)
     \sin\Theta,
     \nonumber\\
&& \{C_8[ij]\} = -\frac{m_W Y_\mu \beta}{4\sqrt{s}c^4_W s^4_W}
      s_\beta\, O_{1k} D_Z\, \real [D_{H_k}]\, 
      \real \left(V_{k;ij}\, Q^{Z*}_{ij}\right)\sin\Theta,
      \nonumber\\
&& [\, C_8[12]] = +\frac{m_W Y_\mu \beta}{4\sqrt{s}c^4_W s^4_W}
      s_\beta\, O_{1k} D_Z\, \imag [D_{H_k}]\, 
      \imag \left(V_{k;12}\, Q^{Z*}_{12}\right)\sin\Theta,
      \nonumber\\
&& \{C_9[ij]\} = +\frac{m_W Y_\mu \beta}{4\sqrt{s}c^4_W s^4_W}
      s_\beta\, O_{1k} D_Z\, \imag [D_{H_k}]\, 
      \real \left(V_{k;ij}\, Q^{Z*}_{ij}\right)\sin\Theta,
      \nonumber\\
&& [\, C_9[12]] = +\frac{m_W Y_\mu \beta}{4\sqrt{s}c^4_W s^4_W}
      s_\beta\, O_{1k} D_Z\, \real [D_{H_k}]\, 
      \imag \left(V_{k;12}\, Q^{Z*}_{12}\right)\sin\Theta,
      \nonumber\\
&& \{C_{10}[ij]\} = +\frac{m_W Y_\mu \beta}{\sqrt{s}c^2_W s^2_W}
      s_\beta\, O_{1k} \real [D_{H_k}]\,\real \left(V_{k;ij}\delta_{ij}
      +D_Z \frac{s^2_W-1/4}{c^2_W s^2_W} V_{k;ij}\,Q^{Z*}_{ij}\right)
      \sin\Theta,
      \nonumber\\
&& [\, C_{10}[12]] = -\frac{m_W Y_\mu \beta}{\sqrt{s}c^4_W s^4_W}
      (s^2_W-1/4) s_\beta O_{1k} D_Z\, \imag [D_{H_k}]\,
      \imag \left(V_{k;12}\,Q^{Z*}_{12}\right)
      \sin\Theta,
      \nonumber\\
&& \{C_{11}[ij]\} = +\frac{m_W Y_\mu \beta}{\sqrt{s}c^2_W s^2_W}
      O_{2k} \real [D_{H_k}]\,\real \left(V_{k;ij}\delta_{ij}
      +D_Z \frac{s^2_W-1/4}{c^2_W s^2_W} V_{k;ij}\,Q^{Z*}_{ij}\right)
      \sin\Theta,
      \nonumber\\
&& [\, C_{11}[12]] = -\frac{m_W Y_\mu \beta}{\sqrt{s}c^4_W s^4_W}
      (s^2_W-1/4) O_{2k} D_Z\, \imag [D_{H_k}]\,
      \imag \left(V_{k;12}\,Q^{Z*}_{ij}\right)
      \sin\Theta,
      \nonumber\\
&& \{C_{12}[ij]\} = -\frac{m_W Y_\mu \beta}{4\sqrt{s}c^4_W s^4_W}
      O_{2k} D_Z \imag [D_{H_k}]\, 
      \real \left(V_{k;ij}\, Q^{Z*}_{ij}\right)\sin\Theta,
      \nonumber\\
&& [\, C_{12}[12]] = -\frac{m_W Y_\mu \beta}{4\sqrt{s}c^4_W s^4_W}
      O_{2k} D_Z \real [D_{H_k}]\, 
      \imag \left(V_{k;12}\, Q^{Z*}_{ij}\right)\sin\Theta.
\label{eq:SV correlations}
\end{eqnarray}
The CP and CP$\tilde{\rm T}$ properties of the observables in
Eq.~(\ref{eq:SV correlations}) can be identified by noting the
following aspects: (i) observables involving $\real [D_{H_k}]$ ($\imag
[D_{H_k}]$) are CP$\tilde{\rm T}$--even (CP$\tilde{\rm T}$--odd). (ii)
The matrix elements $O_{2k}$ and $O_{1k}$ are the CP--even and CP--odd
component of the neutral Higgs boson $H_k$, respectively.  (iii)
$\real \left (V_{k;ij}\,\delta_{ij}\right)$ and $\real \left
(V_{k;ij}\, Q^{Z*}_{ij}\right)$ are CP--even, while $\imag \left
(V_{k;12}\, Q^{Z*}_{12}\right)$ is CP--odd.  For instance,
$\{C_{5}[ij]\}$ is CP--even and CP$\tilde{\rm T}$--even, while $[\,
C_6[12]]$ is CP--odd and CP$\tilde{\rm T}$--even. Of course, this
determination of the CP and CP$\tilde{\rm T}$ properties agrees with
the results of Table~2.

\subsection{Extracting the polarization and rate asymmetries}

In this section we investigate efficient procedures for extracting the
relevant SUSY parameters from the polarized cross sections.  In actual
experiments, a careful analysis of statistical and systematic
uncertainties will be required when extracting the rate and
polarization asymmetries. We do not attempt to estimate realistic
uncertainties based on event simulation or parameter fitting. Instead,
we are interested in the question which of the numerous quantities
listed in the previous Subsection are most sensitive to the
fundamental parameters of the theory, our main focus being on CP--odd
phases in the interaction Lagrangian. 

The conceptually simplest asymmetry is the total unpolarized rate
asymmetry, defined by
\begin{equation} \label{eq:poltot}
{\cal}A_R^{\rm tot} = \frac{ \int \, d\cos\Theta \left( [C_1[12]] +
[C_3[12]] \right) } { \int \, d\cos\Theta \left(\{C_1[12]\} + \{C_3[12]\}
\right)}.
\end{equation}
Unfortunately this asymmetry is usually very small even if CP is
violated in both the Higgs and $\tilde\tau$ sector. We saw in
Eq.~(\ref{eq:zero1}) that $[C_1[12]]$ vanishes in the limit $\Gamma_Z
\rightarrow 0$. Moreover, Eq.~(\ref{eq:sscorr}) shows that $[C_3[12]]$
is proportional to $O_{2k} O_{2l} + \sin^2 \beta O_{1k} O_{1l}$. The
contribution from Higgs exchange can only compete with the gauge boson
exchange contributions if $\sqrt{s} \simeq m_{\rm Higgs}$. Given LEP
search limits for $\tilde\tau_1$, real $\tilde\tau$ pair production
will only be possible at energies $s \gg m_Z^2$.\footnote{This also
implies that in the MSSM the lightest Higgs boson mass must be smaller
than the mass of the lightest tau slepton pair, so that its
contribution to any $\tilde\tau$ pair production process is
negligible.} These two requirements together imply that we need $m_A^2
\gg m_Z^2$. In this ``decoupling limit'' the lightest neutral Higgs
boson eigenstate is given by $H_1 \simeq \cos\beta \phi_1 + \sin\beta
\phi_2$; the couplings of this state are very similar to that of the
Higgs boson of the Standard Model. The rotation matrix $O$ of
Eq.~(\ref{eq:hrot}) is then approximately given by:
\begin{eqnarray} \label{eq:O}
O \simeq
\left(\begin{array}{ccc}
0 & \cos \alpha_H & \sin \alpha_H \\
\cos\beta & \sin\alpha_H \sin\beta & -\cos\alpha_H \sin\beta \\
\sin\beta & -\sin\alpha_H \cos\beta & \cos\alpha_H \cos\beta
      \end{array}\right)\, ,
\end{eqnarray}
where $\sin(2\alpha_H) \neq 0$ signals CP--violation in the Higgs
sector. The value of $\alpha_H$ depends in a complicated manner on the
parameters appearing in the squark mass matrix, and on $m_A$. However,
the structure of the rotation matrix (\ref{eq:O}) implies that
\begin{equation} \label{eq:O0}
O_{2k} O_{2l} + \sin^2 \beta O_{1k} O_{1l} \simeq \sin^2\beta
\delta_{lk}, \ \ l,k \in {2,3} .
\end{equation}
On the other hand, the factor ${\cal D}_{kl}$ appearing in the
expression for $[C_3[12]]$ is nonzero only for $k \neq l$. These two
requirements are incompatible, i.e. the total rate asymmetry is
strongly suppressed.\footnote{This corrects an erroneous numerical
result in Ref.~\cite{CD2}.} 

We thus have to make use of asymmetries that are sensitive to the beam
polarization. To this end we introduce rate and polarization
asymmetries with respect to the unpolarized vector boson exchange and
Higgs boson exchange parts, $\{C_1[ij]\}$ and $\{C_3[ij]\}$,
respectively, as follows:
\begin{eqnarray} \label{eq:pol1}
&& {\cal A}_P\left(C_2[\,ij]\right) = \Omega_2\, \frac{
   \int\, d\cos\Theta \,\{C_2[ij]\}}{
   \int\, d\cos\Theta \,\{C_1[ij]\}}\, ,\nonumber\\[2mm]
&& {\cal A}_R\left(C_2[12]\right) = \Omega_2\, \frac{
   \int\, d\cos\Theta \,[\, C_2[12]]}{
   \int\, d\cos\Theta \,\{C_1[12]\}}\, ,\nonumber\\[2mm]
&& {\cal A}_P\left(C_4[\,ij]\right) = \Omega_4\, \frac{
   \int\, d\cos\Theta \,\{C_4[ij]\}}{
   \int\, d\cos\Theta \,\{C_3[ij]\}}\, ,\nonumber\\[2mm]
&& {\cal A}_R\left(C_4[12]\right) = \Omega_4\, \frac{
   \int\, d\cos\Theta \,[\, C_4[12]]}{
   \int\, d\cos\Theta \,\{C_3[12]\}}\, .
\end{eqnarray}
The remaining $12\times 4$ correlations ($n=5-16$) can only be
measured if at least one of the initial beams is transversely
polarized. In this case the proper normalization involves the sum of
the unpolarized Higgs and gauge boson exchange contributions:
\begin{eqnarray} \label{eq:pol2}
&& {\cal A}_P\left(C_n[\,ij]\right) = \Omega_n\, \frac{
   \int\, d\cos\Theta \,\{C_n[ij]\}}{
   \int\, d\cos\Theta \,\left(\{C_1[ij]\}+\{C_3[ij]\}\right)}\, , 
   \nonumber\\[2mm]
&& {\cal A}_R\left(C_n[12]\right) = \Omega_n\, \frac{
   \int\, d\cos\Theta \,[\, C_n[12]]}{
   \int\, d\cos\Theta \,\left(\{C_1[12]\}+\{C_3[12]\}\right)}\, .
\end{eqnarray}
The $\Omega_n$ in Eqs.~(\ref{eq:pol1}) and (\ref{eq:pol2}) are
numerical factors that originate from projecting out the corresponding
observable, by adjusting the (anti--)muon polarization and integrating
the distribution $\Sigma_{ij}$ over the azimuthal angle $\Phi$ with an
appropriate weight function, as described below. 

These factors $\Omega_n$ ($n=2,\,4,\, 5 - 16$) depend sensitively on
the degrees of longitudinal and transverse polarizations of the muon
and anti--muon beams achievable at muon colliders.  For the sake of
discussion, let us assume that both the muon and anti--muon beams are
polarized with perfect degrees of longitudinal or transverse
polarization. From Eq.~(\ref{eq:Sigma}) we can then deduce the
algorithms for extracting our asymmetries, which determine
the $\Omega_n$. In the following discussion we always assume that all
cross sections are appropriately (anti--)symmetrized in the $\tilde 
\tau$ indices.

\begin{itemize}
\item The asymmetries $\propto C_2[ij]$ can be measured by dividing
the difference of cross sections for $P_L = -\bar{P}_L = 1$ and $P_L =
- \bar{P}_L = -1$ by the sum of these cross sections; this gives
$\Omega_2 = 1$.\footnote{Recall that $[C_2[12]]=0$ in the MSSM if the
Z width is neglected.} Note that a CP--transformation leaves the
polarization states used in this prescription invariant. The
asymmetries $\propto C_4[ij]$ can be obtained in the same way from the
cross sections for $P_L = \bar{P}_L = 1$ and $P_L = \bar{P}_L = -1$,
so that $\Omega_4 = 1$ as well. In this case a CP transformation maps
the two polarization states into each other; this difference of cross
sections is therefore a CP--odd quantity, in accord with the results
of Table~1.
\item The asymmetries involving $C_{15}$ and $C_{16}$ can be obtained
by fixing $P_T=\bar{P}_T=1$ and taking the ratio of the difference and
the sum of the distributions for $\cos(\alpha-\bar{\alpha}) =\pm 1$
($\sin(\alpha-\bar{\alpha})=\pm 1$), respectively. (Recall that $P_T =
\bar{P}_T = 1$ implies $P_L = \bar{P}_L = 0$, and that $\alpha -
\bar{\alpha}$ is independent of $\Phi$.) This gives $\Omega_{15} =
\Omega_{16} = 1$. Recall that a CP transformation flips the sign of
$\alpha - \bar{\alpha}$. Since the cosine is invariant under the
change of sign of its argument while the sine changes its sign, the
prescription for extracting $C_{15}$ is CP--even while that for
$C_{16}$ is CP--odd, again in accordance with the CP transformation
properties of the corresponding terms in the squared matrix element;
see Table~1.
\item The asymmetries involving $C_{13}$ and $C_{14}$ (which vanishes
in the MSSM in the limit $\Gamma_Z = 0$) can also be obtained by
chosing $P_T = \bar{P}_T = 1$. However, the angular combinations
$\cos(\alpha+\bar{\alpha})$ ($\sin(\alpha+\bar{\alpha})$) depend on
the azimuthal angle $\Phi$, see Eq.~(\ref{eq:alpha}). Therefore we need
to apply a (CP--even) normalized weight function $\sqrt{2}
\cos(\alpha+\bar{\alpha})$ ($\sqrt{2} \sin(\alpha+\bar{\alpha})$) in
order to extract these asymmetries, where $\alpha + \bar{\alpha} =
\eta - 2 \Phi$. The normalization factors in these weight functions
$f(\Phi)$ are determined by the requirement that $\int d\Phi f(\Phi)$
acts as a projector, which implies $\int_0^{2\pi} d\Phi f^2(\Phi) =
1$. Note that this weight factor should {\em not} be applied in the
denominator of Eq.~(\ref{eq:pol2}). Finally, one has to take the sum of
these weighted integrated cross sections for $\eta = 0$ and $\eta =
\pi$, in order to remove terms $\propto C_{15}$ or $C_{16}$ from the
denominator\footnote{The numerator is independent of $\eta$ after the
weighted integration over $\Phi$.}; note that these two configurations
of transverse polarization vectors are self--conjugate under a CP
transformation, since $\eta = -\pi$ is identical to $\eta = +\pi$. We
thus find $\Omega_{13} = \Omega_{14} = 1/\sqrt{2}$. Since $\alpha +
\bar{\alpha}$ is also invariant under CP--transformations this
extraction prescription is CP--even, in accord with the corresponding
entries in Table~1.
\item The asymmetries $\propto C_n$ ($n=5 - 8$) are obtained from the
cross sections for $P_T=1, \ \bar{P}_T = \bar{P}_L = 0$ and
$\bar{P}_T=1, \ P_T = P_L = 0$, taking $\eta = 0$. $C_5$ and $C_6$ can
be obtained from the (CP--even) sum of the corresponding cross
sections, while $C_7$ and $C_8$ can be determined from the (CP--odd)
difference of cross sections. In addition one needs the weight
function $\sqrt{2} \cos \alpha$ for $C_5$ and $C_7$, and $\sqrt{2}
\sin \alpha$ for $C_6$ and $C_8$; for $\eta=0$ these weight functions
are all CP--even, so that the extraction procedure again has the same
CP property as the corresponding coefficient of the squared matrix
element. This yields $\Omega_{5} = \Omega_{6} = \Omega_{7} =
\Omega_{8} = 1/\sqrt{2}$.
\item The algorithm for extracting the asymmetries $\propto C_n$ ($n=9
- 12$) is complicated. One possible procedure is to measure the
distributions for the polarization combinations $\{P_L=\pm 1,
\bar{P}_T=1\}$ and $\{\bar{P}_L=\pm 1, P_T=1\}$ for $\eta = 0$. Denote
these four distributions by $\Sigma^{(1)}_\pm$ and $\Sigma^{(2)}_\pm$.
A CP--transformation then sends $\Sigma^{(1)}_\pm$ to
$\Sigma^{(2)}_\mp$ and vice versa.  The asymmetries $\propto C_9$ and
$C_{11}$ can then be extracted from $\Sigma^{(1)}_+ - \Sigma^{(1)}_-
\pm \left( \Sigma^{(2)}_+ - \Sigma^{(2)}_- \right)$, using the weight
function $\sqrt{2} \cos\alpha$; note that the plus (minus) sign in
front of the round parentheses gives a CP--odd (CP--even) combination
of cross sections. The asymmetries $\propto C_{10}$ and $\propto
C_{12}$ can be extracted in the same way, if the weight function
$\sqrt{2} \sin\alpha$ is used. This gives $\Omega_9 = \Omega_{10} =
\Omega_{11} = \Omega_{12} = 1/\sqrt{2}$.
\end{itemize} 
In reality the degrees of longitudinal and transverse polarization are
not perfect, in which case the polarization factors $\Omega_n$ should
be multiplied by the relevant (products of) degrees of longitudinal
and transverse polarizations, $\chi_{_L}$ and $\chi_{_T}$
($\bar{\chi}_{_L}$ and $\bar{\chi}_{_T}$), of the incident muon
(anti--muon) beam. As already stated, we will simply take $\chi_{_L} =
\chi_{_T} = \bar{\chi}_{_L} = \bar{\chi}_{_T} = 1$ in the following
numerical analysis.

\section{Numerical results}
\label{sec:result}
\setcounter{footnote}{1}

In this section, we present some numerical analyses of CP violation 
both in the tau--slepton and neutral Higgs boson systems based on a
specific scenario for the relevant SUSY parameters.

The loop--induced CP violation in the Higgs sector can only be large
if both $|\mu|$ and $|A_t|$ (or $|A_b|$, if $\tan\beta\gg 1$) are
sizable \cite{EXCP1,EXCP2,EXCP3,CDL}.  For definiteness we will
present results for $\tan\beta=10$ and the c.m. energy $\sqrt{s}$ near
the two heavy Higgs boson resonances. Since even for $\tan\beta = 10$
the contributions from the bottom (s)quark sector are still quite
small, our results are not sensitive to $m_{\tilde{D}}$ and $A_b$; we
therefore fix $A_b=A_t$ and $m_{\tilde{D}}= m_{\tilde{U}}=
m_{\tilde{Q}}$, although different values for the SU(2) doublet and
singlet soft breaking squark masses, $m_{\tilde{Q}}\neq
m_{\tilde{U}}$, are allowed, and also take equal phases for $A_t$ and
$A_b$.  Since we are basically interested in distinguishing the CP
non--invariant Higgs sector from the CP invariant one, we take for the
re-phasing--invariant phase $\Phi_{A\mu}$ of $A_{t,b}\,\mu\, {\rm
e}^{i\xi}$ two values; 0 (CP invariant case), and $\pi/2$ (almost
maximally CP violating case) while the phase of the gluino mass
$\Phi_{\tilde{g}}$ is set to be zero. We chose the following
``standard set'' of real mass parameters and couplings of the Higgs
and squark sectors:
\begin{eqnarray}
&&  m_{A} = 0.5\,\, {\rm TeV},  \quad\ \
   |A_{t,b}|= 1.0\,\, {\rm TeV},\quad
   |\mu|=2.0\,\,  {\rm TeV},\nonumber\\
&& m_{\tilde{g}}\,= 0.5\,\,  {\rm TeV},\quad 
   \Phi_{\tilde{g}} = 0,\quad
   m_{\tilde{Q}}=m_{\tilde{U}}=m_{\tilde{D}}=1.0\,\,  {\rm TeV}\,.
\label{eq:parameter set}
\end{eqnarray}
In our numerical analysis, the top and bottom quark running masses 
$\bar{m}_t(m_t)= 165$ GeV and $\bar{m}_b (m_b)=4.2$ GeV will be taken.
For $m_A$ much larger than $m_Z$ as in Eq.~(\ref{eq:parameter set}),
two neutral Higgs bosons have almost degenerate masses:
\begin{eqnarray}
\label{eq:higgs masses}
&& \Phi_{A\mu}=\,0  :\ \  m_{H_2}=499.6\, \,
              \, {\rm GeV},\quad
              m_{H_3}=500.0\, \,\, {\rm GeV},\nonumber\\
&& \Phi_{A\mu}=\frac{\pi}{2}  :\ \ m_{H_2}=499.5\,
              \,\, {\rm GeV} ,\quad m_{H_3}=500.4\, \,\, {\rm GeV}.
\end{eqnarray}
In the CP invariant case ($\Phi_{A\mu}=0$) we have $\cos\alpha_H = 0$,
i.e. the state $H_2$ is CP even and the state $H_3$ is CP odd. On the
other hand, our parameter set with $\Phi_{A\mu} = \pi/2$ gives $\cos
\alpha_H = 0.976$, i.e. now the state $H_2$ is mostly (but not
entirely) CP--odd, and $H_3$ is mostly CP--even. Because of the
near--degeneracy of the two heavy Higgs bosons a significant overlap
effect, i.e. interference between these two Higgs boson resonances is
expected in the presence of CP violation. Note that CP violation in
the Higgs sector tends to increase the mass splitting between the two
heavy states; however, in our case it still does not exceed the widths
of the two Higgs bosons, which amount to about 1.2~GeV. 

We work in the general MSSM, i.e. we do not assume unification of
scalar soft breaking parameters (masses and $A-$terms).\footnote{This
implies that we cannot make statements about first or second
generation sfermion masses. Moreover, we do not assume that the
electroweak gauge symmetry is broken radiatively.} The real SUSY
parameters associated with the tau slepton system are thus independent
of the parameters in the Higgs and squark sectors. We take as standard
inputs: 
\begin{eqnarray}
m_{\tilde{\tau}_L}=0.23\, \,{\rm TeV}\,,\quad
m_{\tilde{\tau}_R}=0.18\, \,{\rm TeV}\,,\quad
|A_\tau|=0.5\, {\rm TeV}\,.
\label{eq:stau parameter set}
\end{eqnarray}
The phase $\Phi_{A_\tau}$ of the rephasing invariant quantity $A_\tau
\mu {\rm e}^{i \xi}$ is varied. For reference, we list the tau
slepton masses for $\Phi_{A_\tau}=0, \pi/2$:
\begin{eqnarray}
\label{eq:stau masses}
&& \Phi_{A_\tau}=\,\, 0\,\, :\ \ m_{\tilde{\tau}_1}\,= 82.7\, \,
                       {\rm GeV}, \quad\,
                       m_{\tilde{\tau}_2}= 287.4\, \,{\rm GeV},
        		\nonumber\\
&& \Phi_{A_\tau}=\, \frac{\pi}{2}\,\, :\ \ 
                       m_{\tilde{\tau}_1}= 87.6\, \,{\rm GeV},\,\quad
                       m_{\tilde{\tau}_2}= 286.0\, \,{\rm GeV}.
\end{eqnarray}
Since the c.m. energy is assumed to be around the resonances of the
heavy Higgs bosons, $\sqrt{s} \simeq 500$ GeV, the tau slepton pairs
$\tilde{\tau}^+_1\tilde{\tau}^-_1$ and $\tilde{\tau}^\pm_1
\tilde{\tau}^\mp_2$ can be produced in $\mu^+\mu^-$ collisions, but
heavy ($\tilde{\tau}^+_2 \tilde{\tau}^-_2$) pairs are not accessible,
as can be seen from Eq.~(\ref{eq:higgs masses}).

In the previous section, we have listed 64 polarization observables
which can be constructed if the production plane can be reconstructed
(at least statistically), i.e. if both the polar angle $\Theta$ and
the azimuthal angle $\Phi$ can be measured.  We have found that there
are 10 CP--violating rate asymmetries and 18 CP--violating
polarization asymmetries. For some polarization observables, and for
all rate asymmetries, it is important to experimentally identify the
electric charges of the produced sfermions from their decay
products. These requirements are very difficult to satisfy in the
production of third generation squarks. Moreover, in most SUSY models
squarks are significantly heavier than sleptons, so that squark pair
production may not be possible for $\sqrt{s} \simeq m_A$. This is why
we focus on the production of tau slepton pairs. Even here the
requirement that the heavy Higgs boson masses are larger than the sum
of the tau slepton masses imposes a nontrivial constraint on the
accessible parameter space. Moreover, in the absence of a detailed
Monte Carlo study it is not clear how well the production plane can be
determined experimentally. In the following we therefore consider the
cases without and with direct reconstruction of the production plane
separately.

Before turning to these asymmetries we briefly discuss the behavior of
the total $\tilde\tau^+_1 \tilde\tau_1^-$ and $\tilde\tau_1^+
\tilde\tau_2^-$ cross sections. These total cross sections are not
only necessary to estimate the number of available events for a given
integrated luminosity. Since (apart from overall prefactors) they also
appear in the denominators of Eqs.~(\ref{eq:pol1}) and (\ref{eq:pol2}),
they also significantly affect the $\sqrt{s}-$dependence of the
asymmetries. These total cross sections are shown in
Fig.~\ref{fig:tot}. We see that the $\tilde\tau_1$ pair production
cross section is sizable ($\sim 100$ fb) even away from the Higgs
poles. Since here the dominant contributions come from gauge boson
exchange, there is almost no sensitivity to the CP--violating phase
$\Phi_{A_\tau}$, as discussed earlier. On the other hand, the Higgs
exchange contribution to the total cross section does depend
significantly on this phase. Not only the maximal value of this cross
section, but also the energy where this maximum is reached depends on
the CP--violating phases. This follows from the observation that in
the absence of CP--violation a CP--odd Higgs boson cannot couple to an
identical $\tilde\tau$ pair; however, once CP is violated, both heavy
Higgs bosons contribute. In fact, for $\Phi_{A_\tau} \simeq \pi$ and
$\sqrt{s} \simeq m_A$, Higgs boson exchange clearly dominates the
total $\tilde\tau_1$ pair production cross section.

The dominance of the Higgs boson exchange contributions near the poles
is even more pronounced in the $\tilde\tau_1 \tilde\tau_2$
channel.\footnote{The difference between the total cross sections for
$\tilde\tau_1^+ \tilde\tau_2^-$ and $\tilde\tau_2^+ \tilde\tau_1^-$
production is negligible, as shown in Sec.~4.2.} Our choice
(\ref{eq:stau parameter set}) implies $|\sin 2 \theta_\tau| \simeq 1$,
which maximizes the $Z \tilde\tau_1 \tilde\tau_2$ coupling
(\ref{eq:qz}). Nevertheless the gauge boson exchange contributions are
suppressed compared to the case of $\tilde\tau_1$ pair production by
the absence of photon exchange contributions and by the smaller
available phase space ($\beta_{12}^3 \simeq 0.24$, as compared to
$\beta_{11}^3 \simeq 0.83$). Note that the Higgs boson exchange
contribution only scales like $\beta$, and is hence much less phase
space suppressed. Moreover, even in the absence of CP--violation both
Higgs bosons can contribute to $\tilde\tau_1 \tilde\tau_2$
production. Nevertheless a strong dependence on the value of
$\Phi_{A_\tau}$ can be seen also in this case. Note that the
$\tilde\tau_1 \tilde\tau_2$ production cross section typically changes
by more than a factor of 10 over the shown range of energies, as
compared to a variation by a factor 2 to 5 in case of $\tilde\tau_1$
pair production. The energy dependence of the total cross section is
therefore especially important for the understanding of the $\sqrt{s}$
dependence of asymmetries in the $\tilde\tau_1 \tilde\tau_2$ channel,
which includes all rate asymmetries. Note also that the peak cross
section, which falls in the range of $\sim 50$ to $\sim 300$ fb, is
sizable also in this channel, when compared to the anticipated
luminosity of $\sim 10$ fb$^{-1}$ per year \cite{MUCOL}. Finally, we
note that in the absence of CP--violation in the Higgs sector the
total cross sections depend are even functions of the phase
$\Phi_{A_\tau}$. This explains why the curves for $\Phi_{A_\tau} =
\phi$ coincide with those for $\Phi_{A_\tau} = 2 \pi - \phi$ in
Figs.~\ref{fig:tot}(a) and (c). However, this degeneracy is lifted if
$\Phi_{A\mu} \neq 0$.

\addtocounter{figure}{2}
\begin{center}
\begin{figure}[htb]
\vspace*{-5cm}
\hspace*{-1.0cm}
\epsfxsize=18cm \epsfbox{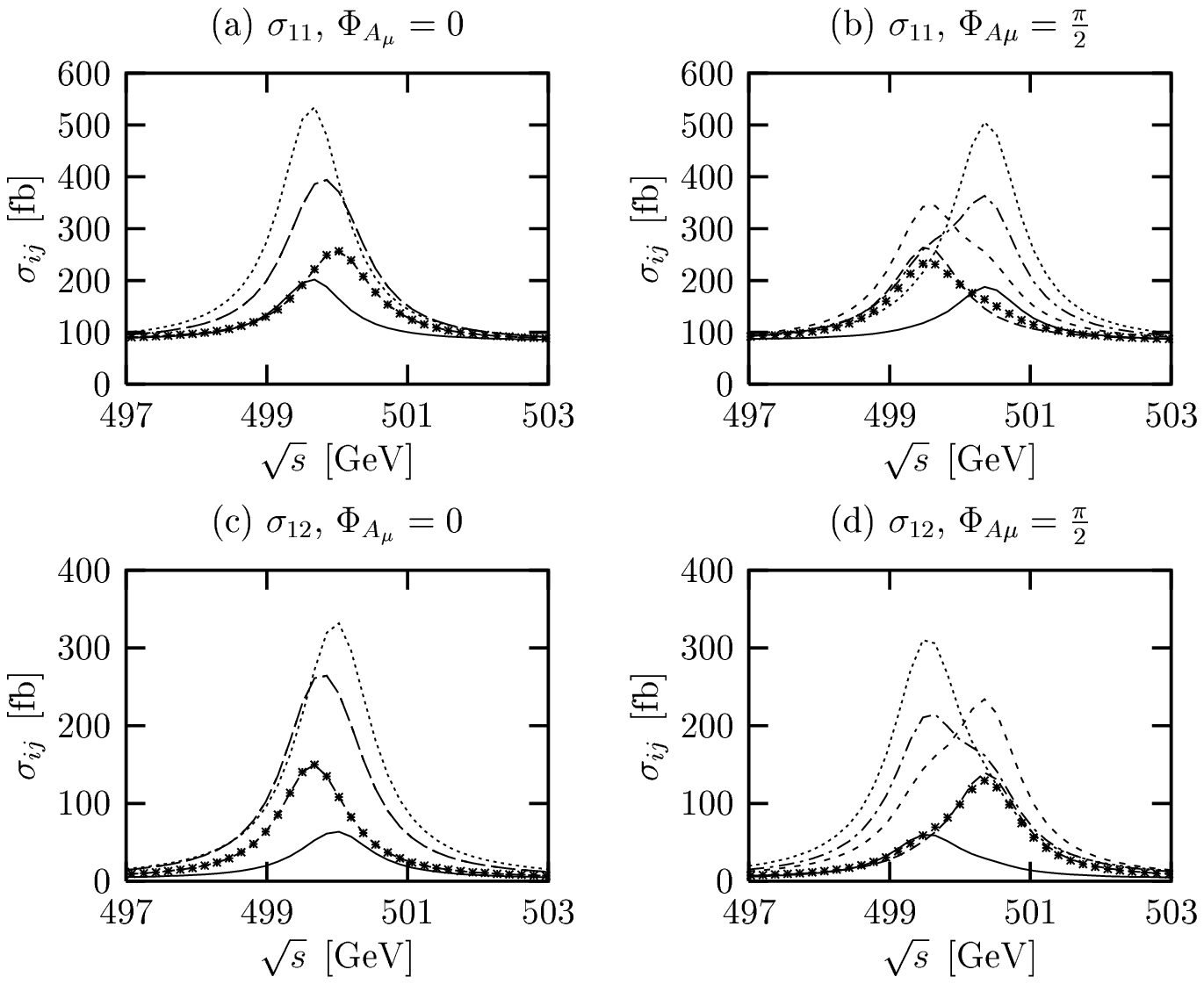}
\vspace*{-12cm}
\caption{\it The $\sqrt{s}$ dependence of the total $\tilde \tau_1$
pair cross section (a,b) and the mixed $\tilde\tau_1^+ \tilde\tau_2^-$
cross section (c,d) near the heavy Higgs boson resonances, with the
SUSY parameter set of Eqs.~(\ref{eq:parameter set}), (\ref{eq:higgs
masses}) and (\ref{eq:stau parameter set}). Frames (a) and (c) are for
$\Phi_{A\mu}=0$, while frames (b) and (d) are for $\Phi_{A\mu}=\pi/2$,
The six lines in each frame are for $\Phi_{A_\tau} = 0$ (solid),
$\pi/3$ (long dashed), $2\pi/3$ (dot--dashed), $\pi$ (dotted),
$4\pi/3$ (short dashed) and $5\pi/3$ (heavily dotted), respectively.}
\vspace*{-0.5cm}
\label{fig:tot}
\end{figure}
\end{center}
%

\subsection{Without direct reconstruction of the production plane}
\setcounter{footnote}{0}

If the production plane is not reconstructed, all the SV correlations
are averaged away when integrating over the azimuthal angle $\Phi$,
but the polarization and rate asymmetries corresponding to the
coefficients $\{C_1,C_2;\, C_3, C_4, C_{15}, C_{16}\}$ survive. As
emphasized earlier, the VV correlations are, however, determined by
the SUSY parameters associated with the tau slepton system without any
CP--preserving phases, if the $Z$ boson width is neglected. In this
case the CP$\tilde{\rm T}$--odd VV rate asymmetries ${\cal
A}_R(C_{1,2}[12])$ vanish so that there exist no non--trivial
CP--violating VV asymmetries\footnote{From now on, let us denote the
asymmetries with SS, VV and SV correlations in the numerators by SS,
VV and SV asymmetries, respectively.} for probing the CP phases in the
tau slepton mass matrix. This immediately implies that $\tilde\tau$
pair production in $e^+e^-$ collisions is of no use for directly
measuring the CP phase associated with the tau slepton
system.\footnote{An exception to this rule can occur if the two
$\tilde\tau$ mass eigenstates happen to be nearly degenerate, with
mass splitting comparable to their decay widths \cite{CD1}.} 

On the other hand, the SS polarization and rate asymmetries are
nonzero. The 6 CP--violating SS polarization asymmetries ${\cal
A}_P(C_4[ij])$ and ${\cal A}_P(C_{16}[ij])$ depend on CP violation in
the $\tilde\tau$ sector only through the CP--even combination of
couplings $\real(V_{k;ij}V^*_{l;ij})$, see Eq.~(\ref{eq:sscorr}); these
asymmetries are CP--odd, since they are also proportional to the
CP--odd combinations $O_{1k} O_{2l} \mp O_{2k} O_{1l}$ of Higgs mixing
angles. The product of this combination of Higgs$- \tilde\tau
\tilde\tau$ couplings and Higgs mixing angles vanishes if CP is
conserved, but in general one can expect a nonzero result in the
presence of CP violation in either the Higgs or $\tilde\tau$
sector. If there is no CP--violation in the Higgs sector ($\Phi_{A\mu}
= 0$), these (and all other CP--violating) asymmetries are odd
functions of $\sin \Phi_{A_\tau}$, i.e. the transformation
$\Phi_{A_\tau} \rightarrow 2 \pi - \Phi_{A_\tau}$ leads to a change of
sign of these asymmetries, leaving their absolute values
unchanged. Since these are polarization asymmetries, they should be
nonzero even for the production of two equal $\tilde\tau$ states, in
particular for $\tilde\tau_1$ pair production. 

\begin{center}
\begin{figure}[htb]
\hspace*{3.0cm}
\epsfxsize=9cm \epsfbox{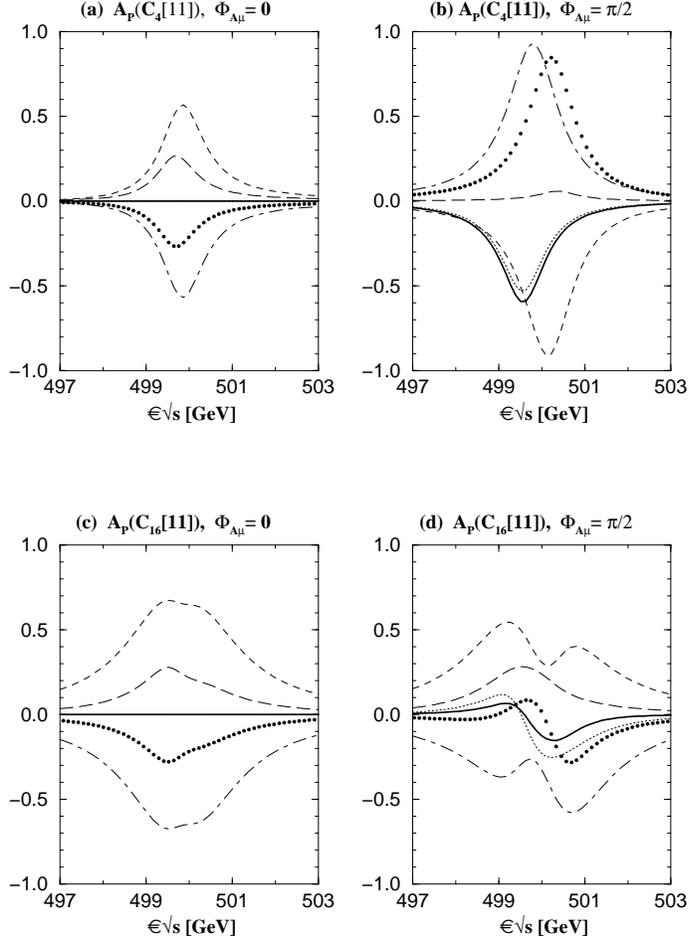}
\caption{\it The $\sqrt{s}$ dependence of the CP--odd SS polarization
             asymmetry ${\cal A}_P(C_4[11])$ for (a) $\Phi_{A\mu}=0$ and (b)
             $\Phi_{A\mu}=\pi/2$, and the CP--odd SS polarization asymmetry 
	     ${\cal A}_P(C_{16}[11])$ for (c) $\Phi_{A\mu}=0$ and 
	     (d) $\Phi_{A\mu}=\pi/2$.  Parameters and notation are as 
	     in Fig.~3.}
\vspace*{-0.5cm}
\label{fig:fig3}
\end{figure}
\end{center}

These expectations are borne out by the numerical results shown in
Fig.~\ref{fig:fig3}. Here the upper two frames show $A_P(C_4[11])$ and
the lower two frames show $A_P(C_{16}[11])$, while the left (right)
frames are for a CP--conserving Higgs sector ($\Phi_{A\mu} =
\pi/2$). Since $A_P(C_4[11])$ is CP$\tilde{\rm T}$--odd, it is
proportional to the imaginary part of the product of relevant Higgs
boson propagators, while $A_P(C_{16})$, being CP$\tilde{\rm T}$--even,
is proportional to the real part of the product of
propagators. Eq.~(\ref{eq:ssprops}) then explains why $A_P(C_4[11])$
decreases much faster as one moves away from the Higgs poles than
$A_P(C_{16}[11])$ does. The results of Fig.~\ref{fig:fig3} show that
both of these polarization asymmetries are quite sensitive to
$A_\tau$. Also, both asymmetries can be large. Finally, we note the
curious fact that in the presence of sizable CP--violation in both the
Higgs and $\tilde\tau$ sectors some CP--odd asymmetries can be
``accidentally'' suppressed; see the long dashed curve (for
$\Phi_{A_\tau} = \pi/3$) in Fig.~\ref{fig:fig3}(b). 

\begin{center}
\begin{figure}[htb]
\vspace*{-10cm}
\hspace*{-5.5cm}
\epsfxsize=40cm \epsfbox{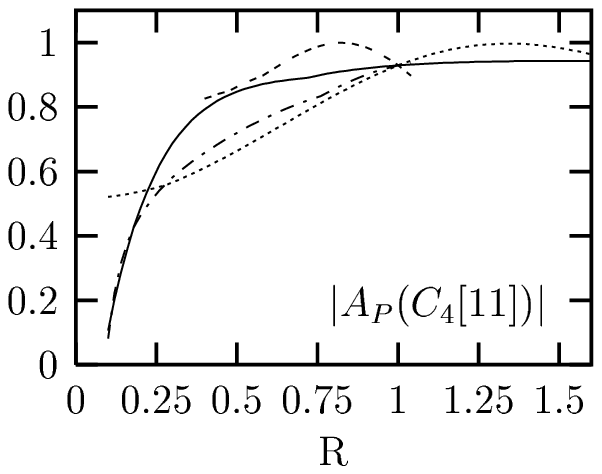}
\vspace*{-36.5cm}
\caption{\it The maximal value of $|{\cal A}_P(C_4[11])|$ is shown for
$\Phi_{A\mu} = \pi/2$ and $\Phi_{A_\tau} = 4 \pi/3$, where ``maximal''
means that the optimal value of $\sqrt{s}$ has been taken. The other
input parameters have been fixed to their ``standard'' values, except
that one of these parameters has been varied along each curve. The
$x-$axis is the ratio of this parameter to its ``standard'' value:
for the solid curve, $R = |A_\tau|/(0.5 \ {\rm TeV})$; for the dashed
curve, $R = \tan\beta/10$; for the dot--dashed curve, $R = |\mu|/(2
{\rm TeV})$; and for the dotted curve, $R = |A_t|/(1 \ {\rm
TeV})$. The upper bounds on $\tan\beta$ and $|\mu|$ come from the
requirement $m_{\tilde{\tau}_1} \geq 80$ GeV, while the lower bound on
$\tan\beta$ comes from the bound $m_{H_1} \geq 111$ GeV.}
\vspace*{-0.5cm}
\label{fig:scan}
\end{figure}
\end{center}

The possibly large size of $A_P(C_4[11])$ is very encouraging, since
this asymmetry can be measured with only longitudinally polarized
beams; this is important, since it might be technically quite
difficult to produce transversely polarized muon beams. Moreover, this
asymmetry is linear in polarization; this means that the precision
with which it can be measured also depends ``only'' linearly on the
achievable degree of polarization. Moreover, in the presence of large
CP--violating phases this asymmetry remains sizable over a large
region of parameter space. This is shown in Fig.~\ref{fig:scan}, which
shows the behaviour of the maximal value of ${\cal A}_P(C_4[11])$ when
one input parameter is moved away from our standard input set
described at the beginning of this Section. The maximum here refers to
the dependence on $\sqrt{s}$, i.e. we plot the asymmetry at the
optimal center--of--mass energy, which depends (slightly) on the input
parameters. This is legitimate, since $\sqrt{s}$ can be freely chosen
in an experiment. In order to show the dependence on different
parameters within a single frame, we chose as $x-$axis the ratio of a
given parameter to its ``standard'' value. The range of variation of
these parameters is restricted by experimental constraints, the most
important ones being the lower bounds $m_{H_1} \geq 111$ GeV (note
that $H_1$ behaves essentially like the Standard Model Higgs boson,
since $m_A^2 \gg M_Z^2$) and $m_{\tilde{\tau}_1} \geq 80$ GeV.

We see that the maximal asymmetry depends only rather weakly on
$\tan\beta$ (dashed curve). However, the total cross section near the
Higgs resonance falls $\propto \tan^4 \beta$ for $\tan^2 \beta \ll
m_t/m_b$, due to the decrease of the couplings of the heavy Higgs
bosons to $\tilde \tau$ sleptons and the simultaneous increase of the
Higgs decay widths into $t \bar t$ quarks. The dependence on
$|A_\tau|$ is also very weak, as long as $|A_\tau| \gsim 200$
GeV. However, sending $|A_\tau| \rightarrow 0$ removes CP--violation
from the $\tilde \tau$ sector and also greatly reduces the couplings
of Higgs bosons to $\tilde \tau$ pairs, see Eq.(\ref{eq:hstaucoup});
hence all CP--odd asymmetries become very small in this limit. The
dependence on $|\mu|$ (dot--dashed curve) is somewhat stronger than
that on $|A_\tau|$, however, we again find a maximal asymmetry $\gsim
0.5$ unless $|\mu|$ is reduced by more than a factor of 3 from its
standard input value of 2 TeV. Finally, the maximal asymmetry remains
finite even as $|A_t| = |A_b| \rightarrow 0$, as shown by the dotted
curve. In this limit there is no CP--violation in the Higgs sector,
but we already saw in Fig.~\ref{fig:fig3} that CP--violation in the
$\tilde \tau$ sector by itself is sufficient to produce a large value
of this asymmetry.

Unfortunately studies of $\tilde\tau_1$ pair production alone do not
suffice to completely determine the parameters of the $\tilde\tau$
sector. Even under the most favorable circumstances such studies can
only determine the values of four real parameters: the mass
$m_{\tilde\tau_1}$, the mixing angle $\theta_\tau$ (these two
parameters can already be determined from VV correlations), and the
couplings of the two heavy Higgs bosons to $\tilde\tau_1$ pairs. On
the other hand, the $\tilde\tau$ mass matrix depends on 6 real
parameters ($m_{\tilde\tau_L}, m_{\tilde\tau_R}, |A_\tau|, |\mu|,
\tan\beta$ and $\Phi_{A_\tau}$). Moreover, if the mixing between
CP--even and CP--odd Higgs states is small, $C_4[11]$ and $C_{16}[11]$
essentially only depend on the {\em product} of the couplings of the
two heavy Higgs bosons to $\tilde\tau_1$ pairs, since in such a
situation the product of mixing angles $O_{1k} O_{2l}$ will be sizable
only for $k \neq l$. In our numerical example this is trivially true
for $\Phi_{A\mu}=0$, but is also approximately correct for
$\Phi_{A\mu}=\pi/2$, where $H_2$ is predominantly CP--odd while $H_3$
is mostly CP--even. Fortunately the total cross section for
$\tilde\tau_1$ pair production for $\sqrt{s} \simeq m_A$ should depend
on a different combination of Higgs couplings, so that it might still
be possible to determine the couplings of both heavy Higgs states to
$\tilde\tau_1$ pairs.

If mixed $\tilde\tau_1^\pm \tilde \tau_2^\mp$ pairs are accessible,
many more quantities become measurable. We saw above that studies of
$\tilde{\tau}_1$ pair production should allow to determine 4 real
parameters in the $\tilde\tau$ sector. In principle the total
$\tilde{\tau}_1^\pm \tilde{\tau}_2^\mp$ cross section, measured in the
vicinity of the heavy Higgs poles, should then be sufficient to
completely fix the parameters of the $\tilde\tau$ sector. However,
this is only true if we assume that the masses and mixing angles of
the neutral Higgs bosons are known. In addition, we have to assume
that $\tilde\tau$ pair production is indeed described by the MSSM, and
that there are no other diagrams contributing; for example, in the
presence of slepton flavor mixing, $t-$channel diagrams might
contribute \cite{japanese}. Analyses of the new asymmetries that
become available if $\tilde{\tau}_1 \tilde{\tau}_2$ pairs can be
produced can then be used to check the consistency of the framework
followed in this paper. Moreover, (some of) these new asymmetries
should also have quite different systematic uncertainties than the
quantities discussed so far. 

\begin{center}
\begin{figure}[htb]
\vspace*{0.0cm}
\hspace*{2.5cm}
\epsfxsize=11cm \epsfbox{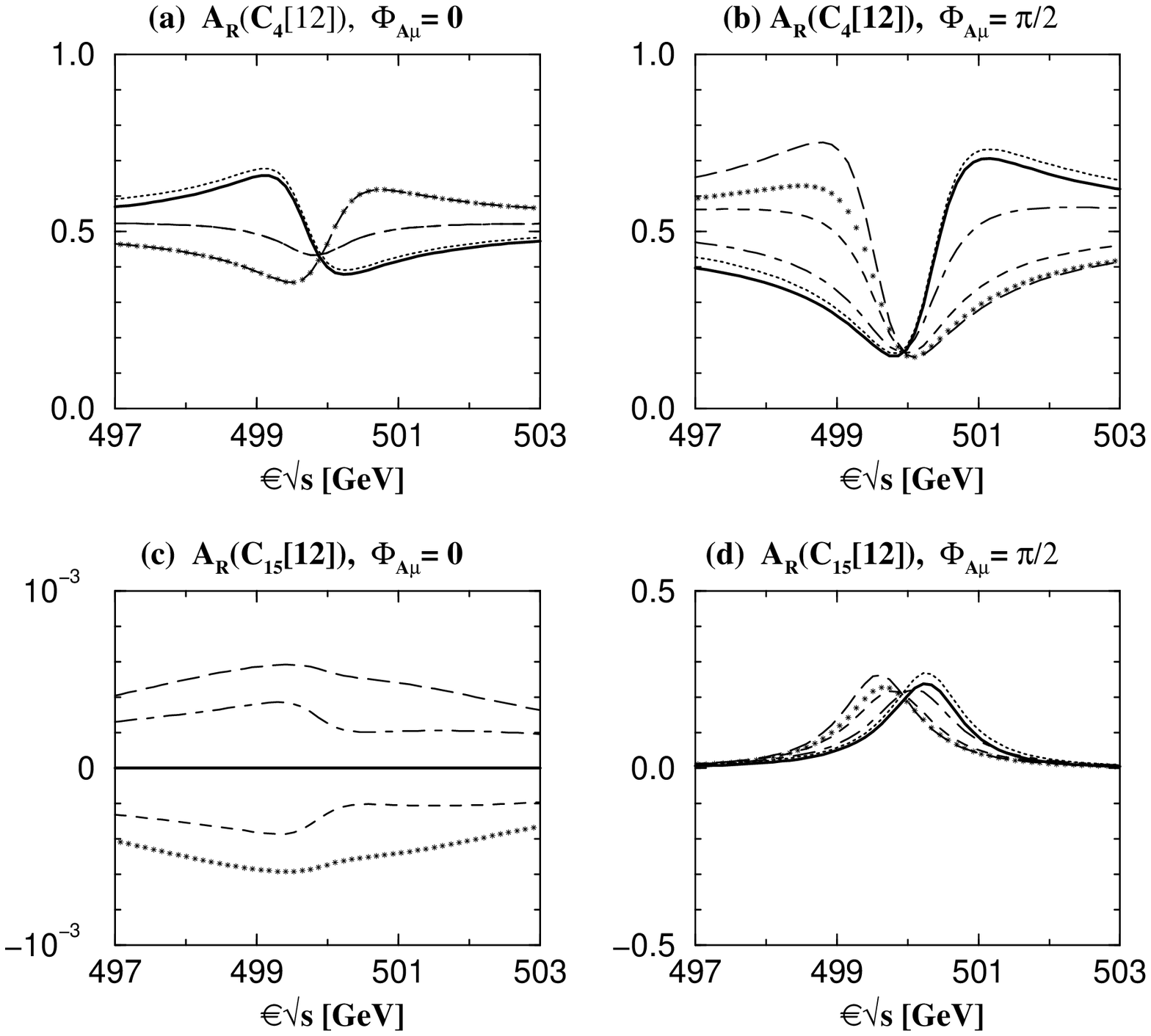}
\caption{\it The $\sqrt{s}$ dependence of the 
             CP--even SS rate asymmetry ${\cal A}_R(C_4[12])$ 
	     for (a) $\Phi_{A}=0$ and (b) $\Phi_{A}=\pi/2$, and 
	     the CP--odd SS rate asymmetry ${\cal A}_R(C_{15}[12])$ 
	     for (c) $\Phi_A=0$ and (d) $\Phi_A=\pi/2$. Parameters and
             notations are as in Fig.~3.}
\vspace*{-0.5cm}
\label{fig:fig4}
\end{figure}
\end{center}

In particular, we now have the opportunity to study rate asymmetries
in addition to polarization asymmetries. All the SS rate asymmetries,
i.e.  the two CP--even asymmetries ${\cal A}_R(C_{4}[12])$ and ${\cal
A}_R(C_{16}[12])$, and the two CP--odd asymmetries ${\cal
A}_R(C_{3}[12])$ and ${\cal A}_R(C_{15}[12])$, are in principle
sensitive to CP violation in the tau slepton system even without
reconstruction of the production plane. However, we already saw at the
beginning of Sec.~4.2 that $[C_3[12]]$ is always very small in the
MSSM with $m_A^2 \gg m_Z^2$. Moreover, we find numerically that
$A_R(C_{16}[12])$ is relatively insensitive to $\Phi_{A_\tau}$, in
particular if there is no CP--violation in the Higgs sector. In
Fig.~\ref{fig:fig4} we therefore show $A_R(C_4[12])$ (top row) and
$A_R(C_{15}[12])$ (bottom row).

The first of these asymmetries is even under both CP and CP$\tilde{\rm
T}$. It is therefore nonzero even if $\Phi_{A\mu} = \Phi_{A_\tau} =
0$, and also remains finite far away from the heavy Higgs poles;
recall that this asymmetry is normalized to the (unpolarized) squared
Higgs boson exchange contribution only, see
Eq.~(\ref{eq:pol1}). However, this is only true for completely
(transversely) polarized beams; and even in this case the number of
events will become very small as one moves away from the Higgs
poles. Nevertheless measuring this asymmetry in the vicinity of the
pole should yield useful information. Note that in the absence of
CP--violation in the Higgs sector, the asymmetry depends only on
$\cos\Phi_{A_\tau}$, i.e. it becomes invariant under $\Phi_{A_\tau}
\rightarrow 2\pi - \Phi_{A_\tau}$; this is true for all other
CP--conserving asymmetries as well. However, this degeneracy is
broken if $\Phi_{A\mu} \neq 0$. Generally we find that CP-violation in
the Higgs sector increases the sensitivity to the rephasing invariant
phase in the $\tilde\tau$ sector. 

This is true also for $A_R(C_{15}[12])$. In fact, this asymmetry is
very small if there is no CP--violation in the Higgs sector. The
reason is that the products of couplings $O_{2k} O_{2l}$ and $O_{1k}
O_{1l}$ appearing in this asymmetry, see Eq.~(\ref{eq:sscorr}), are
almost proportional to $\delta_{kl}$ if the heavy CP--even and CP--odd
Higgs states do not mix, while the combination ${\cal D}_{kl}$ of
Higgs boson propagators vanishes for $k=l$. The interference between
the heavy and light CP--even Higgs states then at best gives an
asymmetry of order $10^{-3}$, which is not measurable with realistic
luminosities. On the other hand, once CP is violated in the Higgs
sector, this asymmetry can become large. However, since it is
CP$\tilde{\rm T}-$odd, it decreases quickly as we move away from the
Higgs poles. Moreover, the results of Fig.~\ref{fig:fig4} indicate
that this asymmetry is not very sensitive to $\Phi_{A_\tau}$ even if
$\Phi_{A\mu} \neq 0$. 

%

\subsection{With direct reconstruction of the production plane}

If the production plane of the tau slepton pair can be reconstructed
efficiently, the SS asymmetries related to $C_{13}$ and $C_{14}$
become accessible. More importantly, all SV correlations can now be
studied. These asymmetries are linear in Higgs propagators, which
implies that they are less sensitive to the strength of the Yukawa
coupling of the muon than the SS correlations are. Moreover, in
principle asymmetries $\propto \Re e[D_{H_k}]$ allow one to ``switch
off'' the contribution from Higgs mass eigenstate $H_k$ simply by
setting $\sqrt{s} = m_{H_k}$, since the real part of the Higgs
propagator vanishes there. In this manner one might be able to cleanly
isolate the contribution from the second heavy Higgs boson. In this
section, based on the parameter sets (\ref{eq:parameter set}) and
(\ref{eq:stau parameter set}), we investigate quantitatively whether
the SV asymmetries can provide us with additional useful information
on the $\tilde\tau$ sector. 

In principle SV correlations can already be studied in polarization
asymmetries in the $\tilde\tau_1^+ \tilde\tau_1^-$
channel. Unfortunately we find numerically that these asymmetries are
always rather small, with absolute value $\lsim 0.2$. This is true
also for polarization asymmetries in the mixed $\tilde\tau_1^\pm
\tilde\tau_2^\mp$ channel. We are thus left with the 8 rate
asymmetries ${\cal A}_R(C_n[12])$ $(n=5$ - $12$). Eq.~(\ref{eq:SV
correlations}) shows that all these asymmetries depend on the
parameters of the $\tilde\tau$ sector through the combination of
couplings
\begin{equation} \label{eq:ik}
I_k \equiv \Im m \left(V_{k;12} Q^{Z\ast}_{12} \right),
\end{equation}
as well as through the masses of the $\tilde\tau$ sleptons. A closer
look at the explicit expressions for these rate asymmetries shows that
the four asymmetries ${\cal A}_R(C_n[12])$ ($n=6,\,7,\,10,\,11$) are
proportional to the small numerical factor $4s^2_W-1\approx -0.07$ for
$s^2_W\approx 0.233$; this comes from the vector coupling of the $Z$
boson to muons. We will therefore not consider these asymmetries, and
instead concentrate on the remaining four asymmetries, which stem from
the axial vector coupling of the $Z$ boson to muons. They satisfy the
proportionality relations:
\begin{eqnarray}
&& {\cal A}_R(C_5[12])\,\, \propto -O_{2k}\, \imag [D_{H_k}]\, I_k,
     \nonumber\\
&& {\cal A}_R(C_8[12])\,\, \propto +s_\beta\,O_{1k}\,\imag [D_{H_k}]\, 
     I_k,
     \nonumber\\
&& {\cal A}_R(C_9[12])\,\, \propto +s_\beta\,O_{1k}\,\real [D_{H_k}]\, 
     I_k,
     \nonumber\\
&& {\cal A}_R(C_{12}[12]) \propto -O_{2k}\, \real [D_{H_k}]\, 
     I_k.
\label{eq:SV rate asymmetries} 
\end{eqnarray}
\begin{center}
\begin{figure}[htb]
\vspace*{0.0cm}
\hspace*{2.5cm}
\epsfxsize=11cm \epsfbox{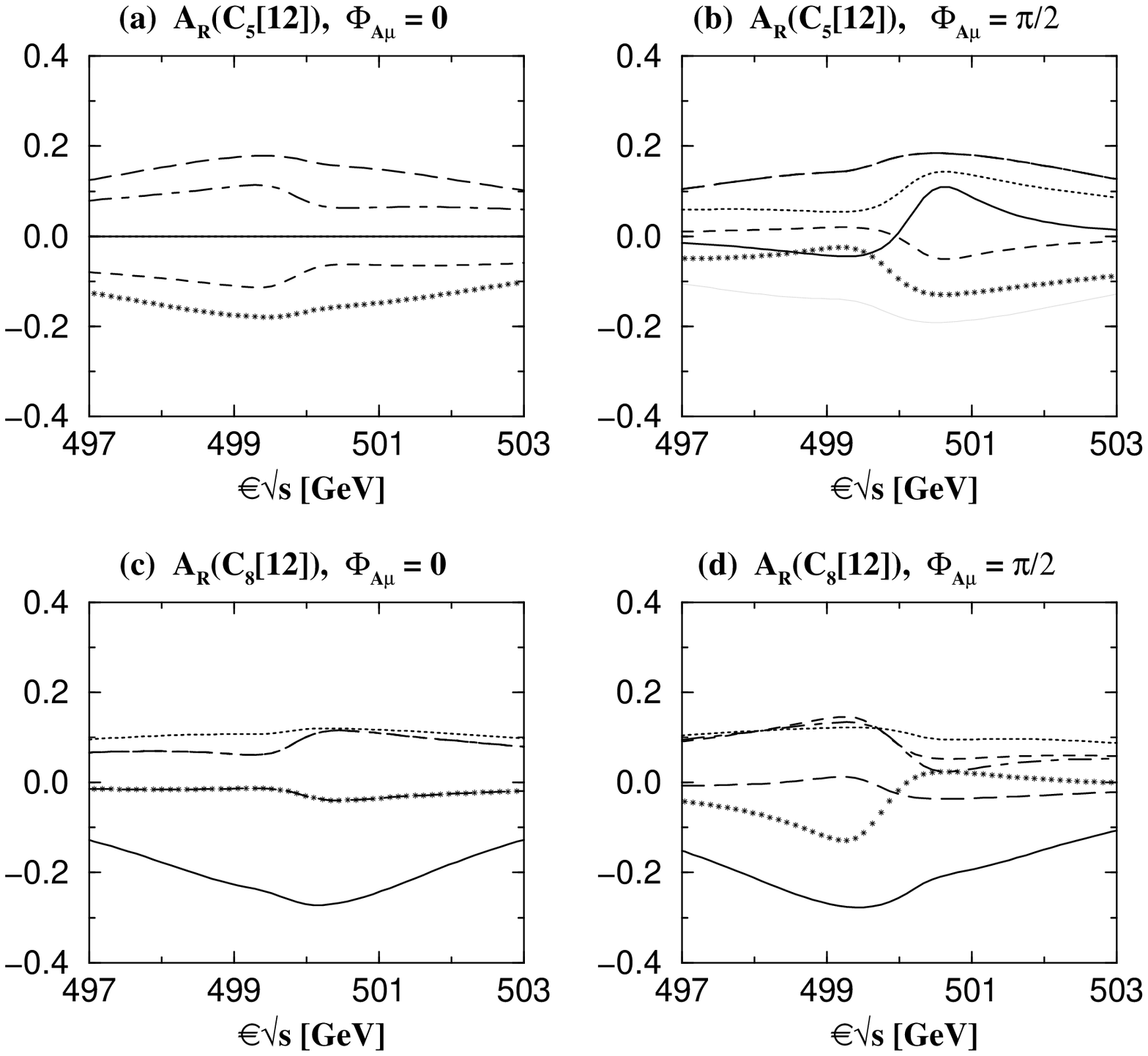}
\caption{\it The $\sqrt{s}$ dependence of the CP--violating SV rate asymmetry 
             ${\cal A}_R(C_5[12])$ for (a) $\Phi_{A\mu}=0$ and (b) 
             $\Phi_{A\mu}=\pi/2$, and the CP--conserving SV rate asymmetry 
             ${\cal A}_R(C_8[12])$ for (c) $\Phi_{A\mu}=0$ and 
	     (d) $\Phi_{A\mu}=\pi/2$, respectively. Notations and
             parameters are as in Fig.~\ref{fig:tot}.}
\vspace*{-0.5cm}
\label{fig:fig5}
\end{figure}
\end{center}
These relations show that the CP--conserving asymmetries ${\cal
A}_R(C_{8,9})$ involve the CP--odd components ($O_{1k}$) of the Higgs
bosons, while the CP--violating asymmetries ${\cal A}_R(C_{5,12})$
involve the CP--even components ($O_{2k}$) of the Higgs bosons. This
means that in the absence of CP violation in the Higgs sector, only
one of the two heavy Higgs bosons contributes to a given asymmetry. In
our case this remains approximately true even if $\Phi_{A\mu} \neq 0$,
since, as noted earlier, the quantity $\sin(2 \alpha_H)$, which
measures the strength of CP--violation in the Higgs sector, is quite
small. In other words, in our examples one does not even have to tune
$\sqrt{s} \simeq m_{H_k}$ in order to isolate the contribution of a
specific Higgs boson.

Note also that the two CP$\tilde{\rm T}$--odd asymmetries ${\cal
A}_R(C_{5,8}[12])$ have their peaks at the heavy Higgs boson poles,
but they are suppressed far away from those poles due to $\imag
[D_{H_k}]$. However, Fig.~\ref{fig:fig5} shows that these asymmetries
only decrease relatively slowly as one moves away from the poles. This
can be understood from the strong energy dependence of the total
$\tilde\tau_1 \tilde\tau_2$ production cross section shown in
Fig.~\ref{fig:tot}: the denominator of ${\cal A}_R(C_{5,8}[12])$ also
decreases quickly as one moves away from the poles.

\begin{center}
\begin{figure}[htb]
\vspace*{0.3cm}
\hspace*{2.5cm}
\epsfxsize=11cm \epsfbox{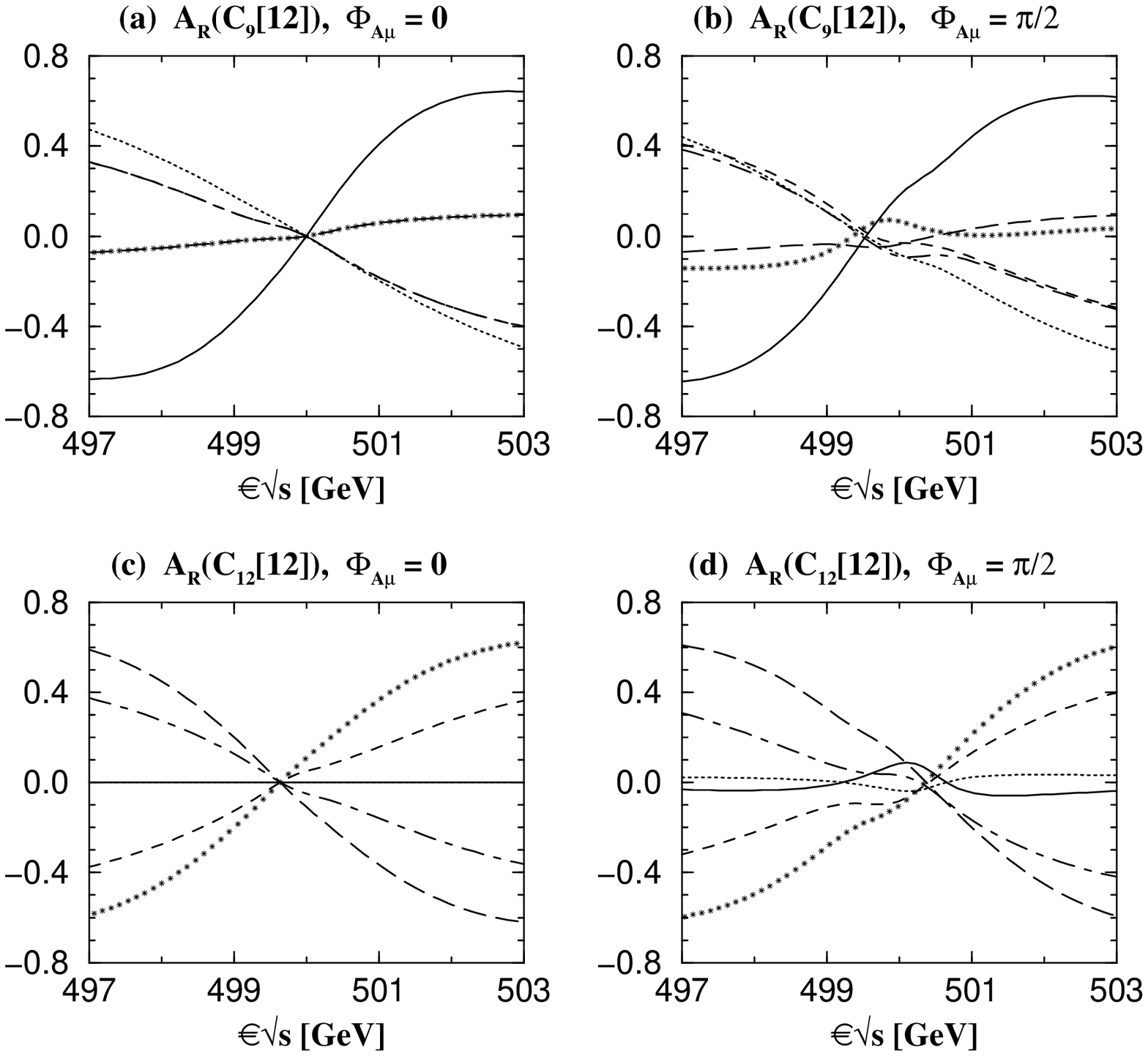}
\caption{\it The $\sqrt{s}$ dependence of the CP--conserving SV rate asymmetry 
             ${\cal A}_R(C_9[12])$ for (a) $\Phi_{A\mu}=0$ and (b) 
             $\Phi_{A\mu}=\pi/2$, and the CP--violating SS rate asymmetry 
             ${\cal A}_R(C_{12}[12])$ for (c) $\Phi_{A\mu}=0$ and 
	     (d) $\Phi_{A\mu}=\pi/2$, respectively. Parameters and
notation are as in Fig.~\ref{fig:tot}.}
\label{fig:fig6}
\end{figure}
\end{center}

On the contrary, Fig.~\ref{fig:fig6} shows that the other two
asymmetries ${\cal A}_R(C_{9,12}[12])$, which are CP$\tilde{\rm
T}$--even, reach their extrema several GeV away from the Higgs
poles. The numerators of these asymmetries actually reach their
extrema at $\sqrt{s} = m_H \pm \Gamma_H/2$, where $m_H$ and $\Gamma_H$
are a typical heavy Higgs mass and decay width, respectively. However,
the rapid decrease of the denominator pushes the extrema further away
from the Higgs poles. Note also that these asymmetries can attain
larger values than the CP$\tilde{\rm T}$--odd asymmetries shown in
Fig.~\ref{fig:fig5}. However, while ${\cal A}_R(C_{5,8}[12])$ can be
measured if only one initial beam is (transversely) polarized, ${\cal
A}_R(C_{9,12}[12])$ can only be determined if both muon beams are
polarized, one having a transverse and the other a longitudinal
polarization. Together with the requirement that the event plane has
to be reconstructed this means that measuring the asymmetries shown in
Fig.~\ref{fig:fig6} will probably pose the biggest challenge to both
collider and detector.\\

\section{Summary and Conclusions}
\label{sec:conclusion}
\setcounter{footnote}{1}

We have performed a detailed, systematic investigation of the signals
for CP violation in the neutral Higgs boson and tau--slepton systems
through the production of tau slepton pairs in polarized $\mu^+\mu^-$
collisions, $\mu^-\mu^+\rightarrow\tilde{\tau}_i^- \tilde{\tau}_j^+$
with the labels $i,j = 1,2$ for the two $\tau$ slepton mass
eigenstates. We worked in the framework of the MSSM with exact
R--parity and negligible flavor mixing. The relevant sources of CP
violation can then be found in soft breaking terms associated with
third generation sfermions, as well as the $\mu-$parameter. CP
violation in the $\tilde \tau$ sector contributes to CP--odd
asymmetries at the tree--level.  CP violation in the $\tilde t$ and/or
$\tilde b$ sector leads to mixing between CP--even and CP--odd Higgs
current eigenstates. Even though this CP--violating Higgs mixing only
proceeds through loop diagrams, it can give rise to ${\cal O}(1)$
CP--violating asymmetries even in the absence of other CP phases.

The expression for the cross section for $\tilde \tau$ pair production
from a $\mu^+ \mu^-$ initial state with arbitrary (possibly
transverse) polarization contains 16 terms. We have classified the
behavior of these terms under CP and CP$\tilde{\rm T}$
transformations. Terms with CP--even polarization factors can
contribute to CP--violating rate asymmetries, i.e. differences between
the cross sections for $\tilde \tau_1^+ \tilde \tau_2^-$ and $\tilde
\tau_1^- \tilde \tau_2^+$ production. On the other hand, terms with
CP--odd polarization factors contribute to CP--violating polarization
asymmetries that can be probed already in $\tilde \tau_1$ pair
production. In some cases it can also be interesting to construct
asymmetries from these terms which are even under a CP
transformation. These asymmetries are nonzero even in the absence of
CP violation, but they can nevertheless help to determine the to date
unknown parameters in the problem, including phases.

It is reasonable to assume that all the properties of the neutral
Higgs bosons are determined beforehand, e.g. through the study of
$\mu^+ \mu^- \rightarrow f \bar{f}$ production near the Higgs poles,
where $f = \tau, b$ or $t$. The remaining task is thus the
determination of the parameters in the $\tilde \tau$ sector. As a
first step one might want to determine the masses of the accessible
$\tilde \tau$ states, as well as the $\tilde \tau_L - \tilde \tau_R$
mixing angle $\theta_\tau$ through the study of $\tilde \tau$ pair
production away from the Higgs poles. Here the cross section will be
dominated by gauge interactions, and is thus completely determined by
these parameters. The mixing angle $\theta_\tau$ also affects the
polarization of the $\tau$ leptons produced in $\tilde \tau$ decays
\cite{nojiri}. Of course, these measurements can also be performed at
an $e^+e^-$ collider.

\begin{center}
\begin{figure}[htb]
\vspace*{-7.5cm}
\hspace*{-1.0cm}
\epsfxsize=18cm \epsfbox{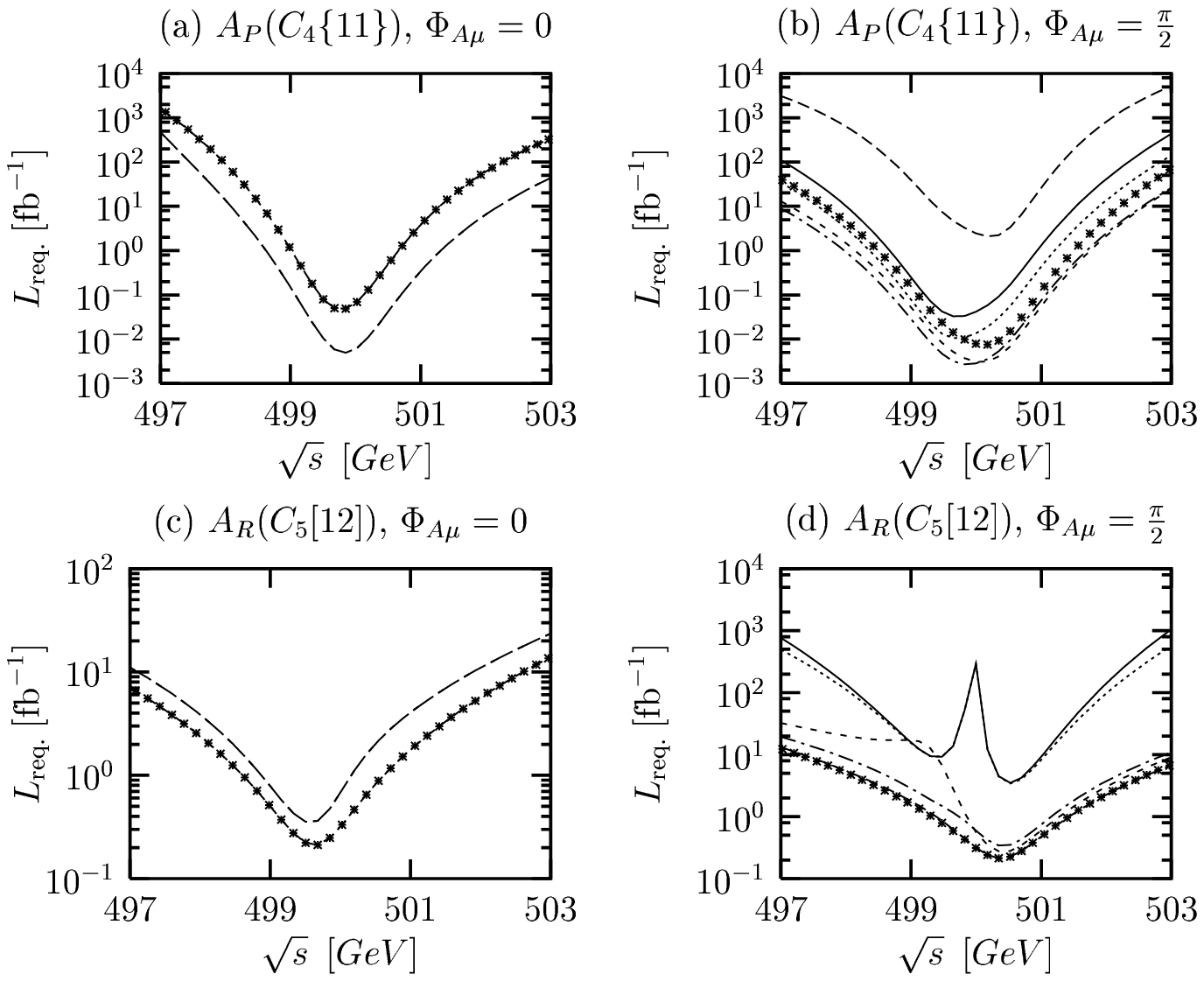}
\vspace*{-9cm}
\caption{\it The $\sqrt{s}$ dependence of the integrated luminosity
         required to detect the asymmetries ${\cal A}_P(C_4[11])$ (a,b) and
         ${\cal A}_R(C_5[12])$ (c,d) with statistical significance of one
         standard deviation, for ideal beam polarization. The luminosity
         increases quadratically with the required number of standard
         deviations, and with the inverse of the degree of beam
         polarization. Parameters and notation are as in Fig.~\ref{fig:tot}.}
\label{fig:signi}
\end{figure}
\end{center}

\vskip -0.9cm

$\tilde \tau$ pair production through gauge interactions can thus at
best give three relations among the six real parameters appearing in
the $\tilde \tau$ mass matrix. The remaining parameters can be
determined if there are sizable Higgs exchange contributions. In
particular, the longitudinal polarization asymmetry ${\cal
A}_P(C_4[11])$ is usually ideally suited for directly probing
CP--violation in the $\tilde \tau$ sector. One does not need to
reconstruct the event plane to measure this asymmetry, nor are
transversely polarized beams required. Moreover, this asymmetry is
linear in the achievable polarization; this is of some importance,
since the actual beam polarization might be significantly less than
the ideal value of 100\% assumed in the numerical results of
Sec.~5. Of course, we have to assume that $\tilde \tau$ pair
production is kinematically allowed in the vicinity of the Higgs
poles. Moreover, the ratio of vevs $\tan\beta$ must not be too small,
since the couplings of both muons and tau (s)leptons to Higgs bosons
scale essentially $\propto \tan\beta$. However, scenarios with very
small $\tan\beta$ are essentially excluded by LEP Higgs searches
anyway. The upper two frames in Fig.~\ref{fig:signi} show that, if
these conditions are met and the CP--violating phases are not very
small, this asymmetry should indeed be measurable with the foreseen
luminosity of about 10 fb$^{-1}$ per year. Here we show the integrated
luminosity required to see a 1$\sigma$ deviation of this asymmetry
from zero, for ideal beam polarization. The required luminosity scales
quadratically with the required number of standard deviations as well
as with the available polarization. For example, a value of 0.01
fb$^{-1}$ means that with a data set of 10 fb$^{-1}$ and 35\% beam
polarization one can see an 11$\sigma$ effect, i.e. this asymmetry
would be measurable with relative precision of about 9\%. We saw in
Fig.~\ref{fig:scan} that this conclusion should remain true over a
wide region of parameter space.

Note, however, that this asymmetry might be ``accidentally''
suppressed, even though there are large CP--violating phases in both
the $\tilde\tau$ and $\tilde t$ sectors. It is therefore important to
try and measure several independent CP--violating asymmetries.
Moreover, analyses of $\tilde \tau_1$ pair production in the MSSM can
at best determine the couplings of $\tilde \tau_1$ pairs to the two
heavy neutral Higgs bosons. LEP search limits on $m_{\tilde \tau_1}$
together with the bound $m_{H_1} \lsim 130$ GeV imply that the
exchange of the lightest neutral Higgs boson $H_1$ contributes
negligibly to $\tilde \tau$ pair production. Simple parameter counting
then implies that, in the absence of any prior knowledge of the
parameters of the mass matrix, one will have to study $\tilde \tau_1
\tilde \tau_2$ production in order to completely determine the
parameters of the $\tilde \tau$ mass matrix. Note that some of these
parameters ($\tan\beta$ as well as $\mu$) also appear in other terms
in the MSSM Lagrangian. It is thus possible that their values will be
known (with some error) beforehand. Even in that case it would be
important to determine them independently from $\tilde \tau$ pair
production, in order to confirm that the MSSM with the given
assumptions can indeed describe this process. Of course, this will
only be possible if $m_A > m_{\tilde \tau_1} + m_{\tilde \tau_2}$.

Unfortunately we found that CP--violating asymmetries that can be
measured in the mixed $\tilde \tau_1 \tilde \tau_2$ channel are
smaller than ${\cal A}_P(C_4[11])$, at least in the vicinity of the
Higgs poles where the event rate is sizable. This is demonstrated by
the lower two frames of Fig.~\ref{fig:signi}, which shows the
luminosity required to see a 1$\sigma$ deviation in the rate asymmetry
${\cal A}_R(C_5[12])$. In most cases the minimal required luminosity
is at least an order of magnitude larger than for ${\cal
A}_P(C_4[11])$. We saw in Sec.~5 that CP$\tilde{\rm T}-$even
asymmetries can be ${\cal O}(1)$ even in the mixed $\tilde \tau_1
\tilde \tau_2$ channel. However, these asymmetries are proportional to
the real parts of heavy Higgs boson propagators, which vanish near the
poles. As a result, the minimal luminosity required for measuring
these asymmetries (not shown) is similar to that required to measure
${\cal A}_R(C_5[12])$, but would have to be taken at $\sqrt{s} \simeq
m_H \pm \Gamma_H$, where $m_H$ and $\Gamma_H$ are the typical mass and
total decay width of the heavy neutral Higgs bosons. It thus seems
unlikely that studies of mixed $\tilde \tau_1 \tilde \tau_2$
production can contribute significantly to the determination of the
CP--violating phase in the $\tilde \tau$ mass matrix. Fortunately the
determination of all parameters of this mass matrix can be completed
by measuring a single quantity in the $\tilde \tau_1 \tilde \tau_2$
channel, for example the total cross section near the Higgs poles. We
saw in Sec.~5 that this cross section is indeed often sizable, and
should thus be readily measurable. A possible problem here might be
the distinction between $\tilde \tau_1$ pair and $\tilde \tau_1 \tilde
\tau_2$ production. This should be fairly easy in the scenario we
studied in Sec.~5, due to the large $\tilde \tau_2 - \tilde \tau_1$
mass splitting, but could be more problematic if this mass splitting
is small.

We thus see that in the chosen framework transverse beam polarization
is not necessary to determine all the free parameters of the $\tilde
\tau$ system. However, in order to test the model one has to
over--constrain it, i.e. there should be more measurements than
parameters. Also, generalizations of the model, e.g. allowing
R--parity violation and/or slepton flavor mixing, are conceivable. In
that case the squared matrix element can still be written as in
Eq.~(\ref{eq:Sigma}), but there will be additional contributions
\cite{greg} to the expressions (\ref{eq:helicity amplitude}), and
hence to the coefficients $C_n$. It is therefore important to measure
as many different asymmetries and distributions as possible, including
those that can only be accessed with transversely polarized
beams. Only then will it be possible to fully exploit the physics
potential of muon colliders to probe the details of supersymmetric
models through scalar $\tau$ production.

\section*{Acknowledgements}

The work of S.Y.C. was supported by a grant from the Korea Research
Foundation (KRF--2000--015--DS0009). The work of M.D. and B.G. was
supported in part by the Deutsche Forschungsgemeinschaft. The work of
J.S.L. was supported by the Japan Society for the Promotion of Science
(JSPS).


\clearpage
\noindent

\end{document}